\documentclass{elsarticle}
 \usepackage{DejaVuSansMono}
\usepackage[T1]{fontenc}
\usepackage[utf8]{inputenc}
\usepackage{ifthen,figlatex}
\usepackage{longtable}
\usepackage{float}
\usepackage{caption}
\usepackage[labelformat=simple]{subcaption}
\usepackage{dblfloatfix}    
\usepackage{wrapfig}
\usepackage{array,multirow,graphicx}
\usepackage{color,soul}
\usepackage[export]{adjustbox}
\usepackage{xspace}
\usepackage{bm}
\usepackage{amsmath,amssymb}
\usepackage[american, english]{babel}
\usepackage{relsize}
\AtBeginDocument{
\definecolor{pdfurlcolor}{rgb}{0,0,0.6}
\definecolor{pdfcitecolor}{rgb}{0,0.6,0}
\definecolor{pdflinkcolor}{rgb}{0.6,0,0}
\definecolor{light}{gray}{.85}
\definecolor{vlight}{gray}{.95}
}
\usepackage{url} 
\usepackage[normalem]{ulem}
\usepackage{todonotes}
\usepackage[symbol]{footmisc}  
\usepackage{fancyvrb}
\usepackage[colorlinks=false,citecolor=pdfcitecolor,urlcolor=pdfurlcolor,linkcolor=pdflinkcolor,pdfborder={0 0 0}]{hyperref}
\usepackage{color,colortbl}
\definecolor{gray98}{rgb}{0.98,0.98,0.98}
\definecolor{gray20}{rgb}{0.20,0.20,0.20}
\definecolor{gray25}{rgb}{0.25,0.25,0.25}
\definecolor{gray16}{rgb}{0.161,0.161,0.161}
\definecolor{gray60}{rgb}{0.6,0.6,0.6}
\definecolor{gray30}{rgb}{0.3,0.3,0.3}
\definecolor{bgray}{RGB}{248, 248, 248}
\definecolor{amgreen}{RGB}{77, 175, 74}
\definecolor{amblu}{RGB}{55, 126, 184}
\definecolor{amred}{RGB}{228,26,28}
\definecolor{amdove}{RGB}{102,102,122}
\usepackage{xcolor}
\usepackage[procnames]{listings}
\definecolor{colorfuncall}{rgb}{0.6,0,0}

\let\oldtexttt=\texttt
\renewcommand\texttt[1]{\oldtexttt{\smaller[1]{#1}}}
\usepackage[binary-units,group-digits,group-separator={,}]{siunitx}
\DeclareSIUnit\flop{Flop}
\DeclareSIUnit\flops{\flop\per\second}
\newcommand{\Num}[1]{\num[group-separator={,}]{#1}\xspace}
\newcommand{\NSI}[2]{\SI[group-separator={,}]{#1}{#2}\xspace}
\usepackage{tikz}
\usetikzlibrary{arrows,shapes,positioning,shadows,trees,calc}
\usepackage{pgfplots}
\pgfplotsset{compat=1.13}
\usepackage{enumitem}
\setlist[itemize,1]{leftmargin=\dimexpr 26pt-.2in}
\usepackage[mode=buildnew]{standalone}
\usepackage[ruled,vlined,english]{algorithm2e}
\DontPrintSemicolon
  \usepackage{graphicx}
  \usepackage{hyperref}
\date{\today}
\title{}
\hypersetup{
 pdfauthor={Tom Cornebize},
 pdftitle={},
 pdfkeywords={},
 pdfsubject={},
 pdfcreator={Emacs 26.1 (Org mode 9.1.13)}, 
 pdflang={English}}
\begin{document}

\newcommand\myemph[1]{\color{colorfuncall}\textbf{#1}}%

\newcommand\labspace[1][-0cm]{\vspace{#1}}
\renewcommand\O{\ensuremath{\mathcal{O}}\xspace}%

\makeatletter
\newcommand{\removelatexerror}{\let\@latex@error\@gobble}
\makeatother

\let\osubsection=\subsection
\def\subsection#1{\osubsection[]{#1}}

\makeatletter
\def\ps@pprintTitle{%
 \let\@oddhead\@empty
 \let\@evenhead\@empty
 \def\@oddfoot{\centerline{\thepage}}%
 \let\@evenfoot\@oddfoot}
\makeatother

\let\oldcite=\cite
\renewcommand\cite[2][]{~\ifthenelse{\equal{#1}{}}{\oldcite{#2}}{\oldcite[#1]{#2}}\xspace}
\let\oldref=\ref
\def\ref#1{~\oldref{#1}\xspace}
\def\eqref#1{~(\oldref{#1})\xspace}
\def\ie{i.e.,\xspace}
\def\eg{e.g.,\xspace}
\def\etal{~\textit{et al.\xspace}}
\newcommand{\AL}[2][inline]{\todo[caption={},color=green!50,#1]{\scriptsize\sf\textbf{$\to$AL:} #2}}
\newcommand{\TC}[2][inline]{\todo[caption={},color=blue!50,#1]{\scriptsize\sf\textbf{$\to$TOM:} #2}}

\title{Simulation-based Optimization and Sensibility Analysis of MPI Applications: Variability Matters}

\renewcommand{\thefootnote}{\fnsymbol{footnote}}
\author{
    Tom Cornebize, Arnaud Legrand\footnote{Corresponding author at: IMAG, 700 Avenue Centrale, Saint-Martin-d'Hères, France\\
        E-mail address: \href{mailto:arnaud.legrand@imag.fr}{arnaud.legrand@imag.fr}\\
    }
}
\address{
    Univ. Grenoble Alpes, CNRS, Inria, Grenoble INP, LIG, 38000 Grenoble, France\\
      firstname.lastname@inria.fr\\[-1.2cm]
}

\newcommand{\model}[2][]{\ensuremath{\mathcal{M}_{#1}\ifthenelse{\equal{#2}{}}{}{\!-\!}{#2}}\xspace}
\newcommand{\modelp}[2][]{\ensuremath{\mathcal{M'}_{#1}\!\ifthenelse{\equal{#2}{}}{}{\!-\!}{#2}}\xspace}
\newcommand{\noise}[2][]{\ensuremath{\mathcal{N}_{#1}\ifthenelse{\equal{#2}{}}{}{\!-\!}{#2}}\xspace}
\newcommand{\noisep}[2][]{\ensuremath{\mathcal{N'}_{#1}\!\ifthenelse{\equal{#2}{}}{}{\!-\!}{#2}}\xspace}
\newcommand{\norm}{\ensuremath{N}\xspace}
\newcommand{\mcdots}{\ensuremath{\!\cdot\!\cdot\!\cdot\!}\xspace}
\begin{abstract}
Finely tuning MPI applications and understanding the influence of key
parameters (number of processes, granularity, collective operation
algorithms, virtual topology, and process placement) is critical to
obtain good performance on supercomputers.  With the high consumption
of running applications at scale, doing so solely to optimize their
performance is particularly costly. Having
inexpensive but faithful predictions of expected performance could be
a great help for researchers and system administrators. The
methodology we propose decouples the complexity of the platform, which
is captured through statistical models of the performance of its main
components (MPI communications, BLAS operations), from the complexity
of adaptive applications by emulating the application and skipping
regular non-MPI parts of the code. We demonstrate the capability of our method with High-Performance
Linpack (HPL), the benchmark used to rank supercomputers in the
TOP500, which requires careful tuning. We briefly present (1) how the
open-source version of HPL can be slightly modified to allow a fast
emulation on a single commodity server at the scale of a
supercomputer. Then we present (2) an extensive (in)validation study
that compares simulation with real experiments and demonstrates our ability to predict the
performance of HPL within a few percent consistently. This study allows us to
identify the main modeling pitfalls (e.g., spatial and temporal node
variability or network heterogeneity and irregular behavior) that need
to be considered.  Last, we show (3) how our ``surrogate'' allows
studying several subtle HPL parameter optimization problems while
accounting for uncertainty on the platform.

\emph{Keywords:} Simulation, validation, sensibility analysis, SimGrid, HPL
\end{abstract}
\maketitle
\section{Introduction}
\label{sec:org3daeb50}
Today, supercomputers with 100,000~cores and more are common, and several
machines beyond the 1,000,000~cores mark are already in
production. These compute resources are interconnected through
complex non-uniform memory hierarchies and network infrastructures.
This complexity requires careful optimization of application
parameters, such as granularity, process organization, or algorithm
choice, as these have an enormous impact on load distribution and
network usage. Scientific application developers and users often spend
a substantial amount of time and effort running their applications at
different scales solely to tune parameters for optimizing
their performance. Whenever actual performance does not match
expectations, it can be challenging to understand whether the
mismatch originates from application misunderstanding or machine
misconfiguration. Similar difficulties are encountered when
(co-)designing supercomputers for specific applications. A large part
of this tuning work could be simplified if a generic and faithful
performance prediction tool was available. This article presents a
decisive step in this direction.

Several techniques have been proposed to predict the performance of a
given application on a supercomputer. A first approach consists in
building a mathematical performance model (i.e., an analytic formula)
accounting for both platform and application key characteristics. However,
it is rarely accurate, except for elementary applications on
highly regular and well-provisioned platforms, and can thus be merely
used to predict broad trends. A more precise approach consists in capturing
a trace of the application at scale and replaying it using a
simulator. This is an effective approach for capacity planning, but
since the application trace is specific to a given set of parameters
(and even specific to a given run for dynamic applications that
exhibit non-deterministic behaviors due to, e.g., the use of asynchronous collective
operations), it cannot be used to study how application parameters
should be set for optimizing performance. The main difficulty resides in capturing and
modeling the interplay between the application and the platform while
faithfully accounting for their respective complexity.  A promising
approach recently pioneered in several tools\cite{smpi,sstmacro,xsim}
consists in emulating the application in a controlled way so that a
platform simulator governs its execution. Although this
approach's scalability is a primary concern that has already
received lots of attention, the accuracy of the simulation is even more
challenging. It is still an open research question since Engelmann and
Naughton\cite{xsim_network} report, for example, an error ranging from
20\% to 40\% for NPB LU when using 128 ranks.

In a previous publication\cite{cornebize:hal-02096571}, we presented how an application like
HPL can be emulated at a reasonable cost on a single commodity server to
study scenarios similar to qualification runs of supercomputers for
the Top500 ranking\cite{TOP500}. We also showed how to predict the performance
of HPL for a specific set of parameters on a recent cluster (running a
thousand MPI ranks) within a few percent of reality. 
The tuning of HPL is generally performed by skilled engineers but the
way it is done is considered as a sensitive information for vendors
and it is thus not well documented. HPL is
particularly challenging to study because it implements several custom
non-trivial MPI collective communication algorithms to 
overlap communications with computations efficiently. Since it is
a tightly coupled application, it is also expected to be quite
sensitive to platform variability (both spatial and temporal) but to
the best of our knowledge, although it is well known by HPL experts,
it has never been properly quantified. We
demonstrated in our previous
publication\cite{cornebize:hal-02096571} that a careful modeling of
variability was a key ingredient to obtain good predictions and we
study this sensitivity more in details hereafter (Section\ref{sec:whatif}). In this
article, we conduct an extensive validation study. We show that our
approach allows us to consistently predict the real-life performance of
HPL within a few percent regardless of its input parameters, thereby
showing that application parameters can be tuned fully in
simulation. The reason why our approach is particularly effective is two-fold: (1) HPL
control flow is data-independent (up to some micro-variations) and (2)
the bulk of communications consists of large messages. Although
our work focuses on HPL, it could be applied to other similar MPI applications
satisfying these conditions.

Throughout this validation, which spanned over two years,
we also highlight key issues that may arise when modeling the platform
and should be carefully addressed to obtain reliable
predictions. Last, given the sensibility of applications to
computing and communication resource variability, we showcase
how to conduct what-if performance analysis of HPC applications
in a capacity planning context.

This article is organized as follows: Section\ref{sec:con} presents
the main characteristics of the HPL application and provides
information on how the runs are conducted on modern supercomputers.
In Section\ref{sec:smpi}, we briefly present the simulator we used for
this work, SimGrid/SMPI, and the modifications of HPL that were
required to obtain a scalable simulation and some initial validation
results presented in an earlier work\cite{cornebize:hal-02096571}. These results
highlight the importance of modeling both spatial and temporal
variability. In Section\ref{sec:validation}, we compare simulation
results with real experiments through two typical HPL performance
studies that cover a wider range of application
parameters. Section\ref{sec:whatif} presents how our HPL surrogate can
be used to study and possibly optimize the performance of HPL in
the presence of uncertainty on the platform.  Section\ref{sec:relwork}
discusses related work and explains how our approach compares with other approaches.
Section\ref{sec:cl} concludes by discussing future work.
\section{Background on High-Performance Linpack}
\label{sec:orgb8a8bb2}
\label{sec:con}
\label{sec:hpl}

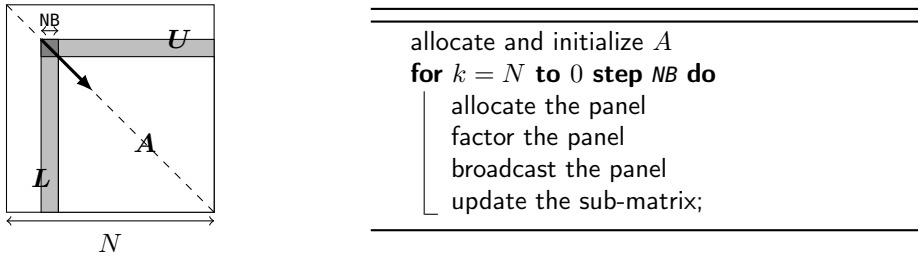
\begin{figure}[t]
  \newcommand{\mykwfn}[1]{{\bf\textsf{#1}}}%
  \SetAlFnt{\sf}%
  \SetKwSty{mykwfn}%
  \SetKw{KwStep}{step}%
  \centering
  \begin{minipage}[m]{0.4\linewidth}
    \begin{tikzpicture}[scale=0.23]
      \draw (0, 0) -- (0, 12) -- (12, 12) -- (12, 0) -- cycle;
      \foreach \i in {2}{
        \draw [fill=lightgray] (\i, 0) -- (\i, 12-\i) -- (12, 12-\i) -- (12, 0) -- cycle;
        \draw [fill=gray] (\i, 12-\i) -- (\i, 12-\i-1) -- (\i+1, 12-\i-1) -- (\i+1, 12-\i) -- cycle;
        \draw[very thick, -latex] (\i,12-\i) -- (\i+2,12-\i-2);
        \draw[<->] (\i, 12-\i+0.5) -- (\i+1, 12-\i+0.5) node [pos=0.5, yshift=+0.15cm] {\scalebox{.8}{\texttt{NB}}};
      }
      \foreach \i in {3}{
        \draw [fill=white] (\i, 0) -- (\i, 12-\i) -- (12, 12-\i) -- (12, 0) -- cycle;
        \draw (\i,12-\i) -- (\i,0);
        \draw[very thick, -latex] (\i,12-\i) -- (\i+2,12-\i-2);
      }
      \draw[dashed] (0, 12) -- (12, 0);
      \node(L) at (2, 2) {\ensuremath{\boldsymbol{L}}};
      \node(U) at (10, 10) {\ensuremath{\boldsymbol{U}}};
      \node(A) at (8, 4) {\ensuremath{\boldsymbol{A}}};
      \draw[<->] (0, -0.5) -- (12, -0.5) node [pos=0.5, yshift=-0.3cm] {$N$};

    \end{tikzpicture}
  \end{minipage}%
  \begin{minipage}[m]{0.6\linewidth}
    \removelatexerror
    \begin{algorithm}[H]
      allocate and initialize $A$\;
      \For{$k=N$ \KwTo $0$ \KwStep \texttt{NB}}{
        allocate the panel\;
        factor the panel\;
        broadcast the panel\;
        update the sub-matrix;
      }
    \end{algorithm}
    \vspace{1em}
  \end{minipage}\vspace{-.5em}
  \caption{Overview of High-Performance Linpack.}\vspace{-1.5em}
  \label{fig:hpl_overview}
\end{figure}

HPL implements a matrix factorization based on a right-looking variant
of the LU factorization with row partial pivoting and allows for multiple
look-ahead depths. In this work, we use the freely-available
reference-implementation of HPL\cite{HPL}, which relies on MPI, and
from which most vendor-specific implementations (e.g., from Intel or
ATOS) have been derived. Figure\ref{fig:hpl_overview} illustrates the
principle of the factorization which consists of a series of panel
factorizations followed by an update of the trailing sub-matrix.  HPL
uses a two-dimensional block-cyclic data distribution of \(A\), which
allows for a smooth load-balancing of the work across iterations.

The sequential computational complexity of this factorization is
\(\mathrm{flop}(N) = \frac{2}{3}N^3 + 2N^2 + \O(N)\) where \(N\) is the
order of the matrix to factorize. The time complexity on a \(P\times Q\)
processor grid can thus be
approximated by $$T(N) \approx \frac{\left(\frac{2}{3}N^3 +
2N^2\right)}{P\cdot{}Q\cdot{}w} + \Theta((P+Q)\cdot{}N^2),$$ where \(w\) is the flop rate of a
single node and the second term corresponds to the communication
overhead which is influenced by the network capacity and many
configuration parameters of
HPL\@.
Indeed, HPL implements several custom
MPI collective communication algorithms to efficiently overlap
communications with computations. The main parameters of HPL are thus:
\begin{itemize}
\item \(N\) is the order of the square matrix \(A\).
\item \texttt{NB} is the ``blocking factor'', \ie the granularity at which HPL
operates when panels are distributed or worked on. This parameter
influences the efficiency of the \texttt{dgemm} BLAS kernel, which is the
kernel used in the sub-matrix updates, but also the efficiency of
MPI communications.
\item \(P\) and \(Q\) denote the number of process rows and process
columns. For this algorithm, the \emph{total} amount of data transfers is
proportional to \((P+Q).N^2\), which generally favors
virtual topologies where \(P\) and \(Q\) are approximately equal.
\item \texttt{RFACT} determines the panel factorization algorithm. Possible values
are \texttt{Crout}, \texttt{left-} or \texttt{right-looking}.
\item \texttt{SWAP} specifies the swapping algorithm used while pivoting. Two
algorithms are available: one is based on a \emph{binary exchange} (along a
virtual tree topology) and the other one is based on a \emph{spread-and-roll}
(with a higher number of parallel communications). HPL also provides
a panel-size threshold triggering a switch from one variant to the
other.
\item \texttt{BCAST} sets the algorithm used to broadcast a panel of columns over
the process columns. Legacy versions of the MPI standard only
supported non-blocking point-to-point communications, which is why
HPL ships with in total 6 self-implemented variants to overlap the
time spent waiting for an incoming panel with updates to the
trailing matrix: \texttt{ring}, \texttt{ring-modified}, \texttt{2-ring}, \texttt{2-ring-modified}, \texttt{long},
and \texttt{long-modified}. The \texttt{modified} versions guarantee that the process
right after the root (\ie the process that will become the root in
the next iteration) receives data first and does not further
participate in the broadcast. This process can thereby start working
on the panel as soon as possible. The \texttt{ring} and \texttt{2-ring} versions each
broadcast along the corresponding virtual topologies while the \texttt{long}
version is a \emph{spread and roll} algorithm where messages are chopped
into \(Q\) pieces. This generally leads to better bandwidth
exploitation. The \texttt{ring} and \texttt{2-ring} variants rely on \texttt{MPI\_Iprobe},
meaning they return control if no message has been fully received
yet, hence facilitating partial overlap of communication with
computations. In HPL 2.1 and 2.2, this capability has been
deactivated for the \texttt{long} and \texttt{long-modified} algorithms. A comment in
the source code states that some machines apparently get stuck when
there are too many ongoing messages.
\item \texttt{DEPTH} controls how many iterations of the outer loop can overlap
with each other. As indicated in the HPL documentation, a depth
equal to 1 often gives better results than a depth equal to 0 for
large problem sizes, but a look-ahead of depth equal to 3 and larger
is not expected to bring any improvement.
\end{itemize}

All the previously listed parameters interact uniquely
with the interconnection network capability and the MPI library to influence
the overall performance of HPL, which makes it very difficult to
predict precisely. To illustrate the diversity of real-life
configurations, we report in Table\ref{fig:typical_run} a few ones
used for the TOP500 ranking that some colleagues agreed to share with
us.

\begin{table}[t]
\vspace{-1em}
\caption{Typical HPL configurations.}
\label{fig:typical_run}
\scalebox{.9}{\begin{tabular}{l|lll}
\multicolumn{1}{l|}{}    & Stampede@TACC             & Theta@ANL                 & \\
\multicolumn{1}{l|}{}    & \#6th \qquad June 2013    & \#18th \qquad Nov. 2017   &  \\
\hline
\texttt{Rpeak}           & \NSI{8520.1}{\tera\flops} & \NSI{9627.2}{\tera\flops} & \\
$N$                      & \Num{3875000}             & \Num{8360352}             & \\
\texttt{NB}              & \Num{1024}                & 336                       & \\
\texttt{P}$\times$\texttt{Q}  & 77$\times$78                   & 32$\times$101                  & \\
\texttt{RFACT}           & Crout                     & Left                      & \\
\texttt{SWAP}            & Binary-exch.              & Binary-exch.              & \\
\texttt{BCAST}           & Long modified             & 2 Ring modified           & \\
\texttt{DEPTH}           & 0                         & 0                         & \\
\hline               
\texttt{Rmax}            & \NSI{5168.1}{\tera\flops} & \NSI{5884.6}{\tera\flops} & \\
Duration                 & 2 hours                   & 28 hours                  & \\
Memory                   & \NSI{120}{\tera\byte}     & \NSI{559}{\tera\byte}     & \\
MPI ranks                & 1/node                    & 1/node                    & \\
\end{tabular}}\vspace{-1em}
\end{table}
\label{sec:con:diff}

The performance typically achieved by supercomputers (\texttt{Rmax}) needs to
be compared to the much larger peak performance (\texttt{Rpeak}).
The difference can be attributed to the node
usage, to the MPI library, to the network topology that
may be unable to deal with the intense communication workload, to
load imbalance among nodes (\eg due to a defect, system noise,\ldots), to the algorithmic structure of
HPL, etc. All these factors make it difficult to know precisely what
performance to expect without running the application at scale.
Due to the complexity of both HPL and the underlying hardware, simple
performance models (analytic expressions based on \(N, P, Q\) and
estimations of platform characteristics as presented in
Section\ref{sec:hpl}) may at best be used to determine broad trends but can by no means
accurately predict the performance for each configuration (\eg consider the
exact effect of HPL's six different broadcast algorithms on network
contention). Additionally, these expressions do not allow
engineers to improve the performance through actively identifying performance bottlenecks.
For complex optimizations such as partially non-blocking
collective communication algorithms intertwined with computations,
a very faithful model of both the application and the platform is
required.
\section{Emulating HPL with SimGrid/SMPI}
\label{sec:orgc7ebd3c}
\label{sec:smpi}
In this section, we present an overview of Simgrid/SMPI and of the
modifications of HPL required to obtain a scalable
simulation and a first validation of the simulations. The results of
Sections\ref{sec:smpi:nutshell}--\ref{sec:validation.single}
previously appeared in a conference
publication\cite{cornebize:hal-02096571} and are included here for
completeness.
\subsection{Simgrid/SMPI in a Nutshell}
\label{sec:orgdee84ec}
\label{sec:smpi:nutshell}
SimGrid\cite{simgrid} is a flexible and open-source simulation
framework that was initially designed in 2000 to study scheduling
heuristics tailored to heterogeneous grid computing environments but
was later extended to study cloud and HPC infrastructures. The
main development goal for SimGrid has been to provide validated
performance models, particularly for scenarios making heavy use of the
network. Such a validation usually consists of comparing simulation
predictions with real experiments to confirm or debunk
and improve network and application models.

SMPI, a simulator based on SimGrid, has been developed and used to
simulate unmodified MPI applications written in C/C++ or
FORTRAN\cite{smpi}. To this end, SMPI maps every MPI rank of the
application onto a lightweight simulation thread. These threads are
then run in mutual exclusion and controlled by SMPI, which measures
the time spent computing between two MPI calls. This duration is
injected in the simulator as a simulated delay, scaled up or down
depending on the speed difference between the simulated machine and
the simulation machine.

The complex optimizations done in real MPI implementations
need to be considered when predicting the performance of
applications.  For instance, the ``eager'' and ``Rendez-vous'' protocols
are selected based on the message size, with each protocol having its
synchronization semantic, which strongly impacts performance.
Another problematic issue is to model network topologies and
contention. SMPI relies on SimGrid's communication models, where each
ongoing communication is represented as a single \emph{flow} (as opposed to a
collection of individual packets). Assuming steady-state,
contention between active communications can then be modeled as a
bandwidth sharing problem while accounting for non-trivial phenomena (\eg
cross-traffic interference\cite{Velho_TOMACS13}).  The time spent in
MPI is thus derived from the SMPI network model that accounts for
MPI peculiarities (depending on the message size), the machine
topology, and the contention with all other ongoing flows. For more
details, we refer the interested reader to\cite{smpi}.

\begin{figure}[ht]
    \centering
    \includegraphics[width=.9\linewidth]{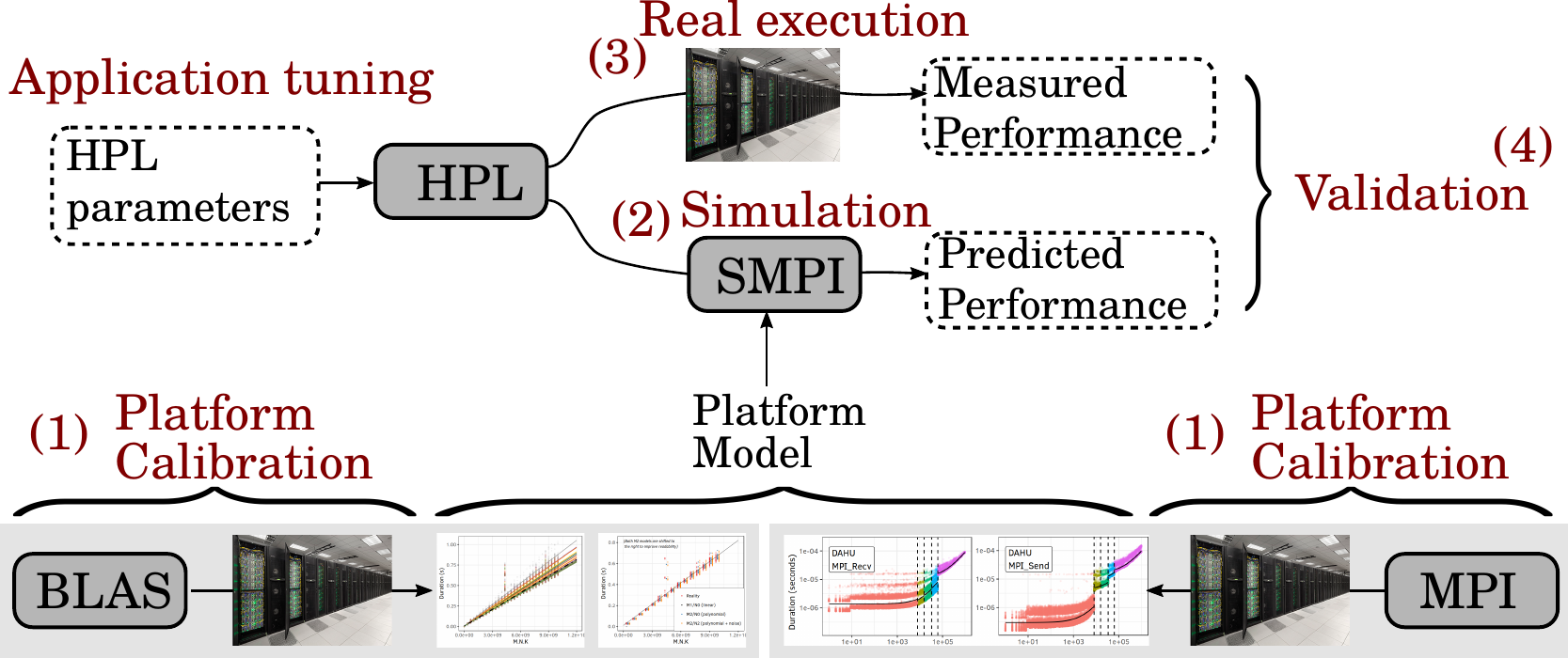}
    \caption{Experimental and simulation workflow with SMPI. }\vspace{-1em}
    \label{fig:smpi_workflow}
    \labspace
\end{figure}

Figure\ref{fig:smpi_workflow} provides an overview of how performance
evaluation studies are conducted with SMPI and in this article. First, a series of
benchmarks is conducted on the target machine to calibrate (step 1) the
platform model. Once this model is built, the application can be
simulated over SMPI (step 2) at low cost to predict performance while varying
its parameters (e.g., for application tuning) without resorting to the target machine anymore. This
approach should be contrasted with the classical one which solely
relies on real executions on the target machine (step 3). In this article, as
we are particularly interested in evaluating how accurate the
predictions are, we propose an extensive comparison of predicted
performance with measured performance (step 4). More precisely, we
show in Sections\ref{sec:validation.single}-\ref{sec:validation} that
predictions are particularly faithful for HPL provided the platform is
calibrated with care (steps 1-4) and we show in Section\ref{sec:whatif} how
specific characteristics of HPL can be studied in simulation by slightly
varying and extrapolating the platform model (steps 1-2).
\subsection{Emulating HPL}
\label{sec:org728a7df}
\label{sec:em}

\begin{figure}[!b]
  \centering
  \lstset{frame=bt,language=C,numbers=none,escapechar=|}
\begin{lstlisting}
#define |\color{colorfuncall}HPL\_dgemm|(layout, TransA, TransB, M, N, K,     \
        alpha, A, lda, B, ldb, beta, C, ldc) ({        \
    double size = ((double)M)*((double)N)*((double)K); \
    double expected_time = 1.029e-11*size + 1.981e-12; \
    |\color{colorfuncall}smpi\_execute\_benched|(expected_time);               \
})
\end{lstlisting}
  \caption{Non-intrusive macro replacement with a very simple performance model.\label{fig:macro_simple}}
\end{figure}
HPL relies heavily on BLAS kernels such as \texttt{dgemm} (for matrix-matrix
multiplication). Since these kernels' output does not influence the
control flow, simulation time can be reduced considerably by
substituting these function calls with a 
performance model of the respective kernel.
Figure\ref{fig:macro_simple} shows an example of this macro-based
mechanism that allows us to keep HPL code modifications to an absolute
minimum. The \texttt{(1.029e-11)} value represents the inverse of the flop rate
for this compute kernel and is obtained by benchmarking the target
nodes. The kernel's estimated duration is calculated based on the given
parameters and passed on to \texttt{smpi\_execute\_benched} that
advances the simulated clock of the executing rank by this estimate.
Skipping compute kernels makes the content of output variables invalid, but in
simulation, only the application's behavior and not the
correctness of computation results are of concern. These minor modifications to the original 
source code  (HPL comprises 16K lines of ANSI C over 149 files, our modifications only
changed 14 files with 286 line insertions and 18 deletions) 
enabled us to simulate the configuration used for the Stampede cluster in
2013 for the TOP500 ranking (see Table\ref{fig:typical_run}) in less
than 62 hours and using \NSI{19}{\giga\byte} on a single node of a commodity
cluster (instead of \NSI{120}{\tera\byte} of RAM over a 6006 node supercomputer). 
Further speed-up could probably be obtained by modifying HPL further,
but our primary interest in this article is on the prediction quality.

\begin{figure*}[!t]
  \vspace{-1em}
  \begin{minipage}[t]{\linewidth}\null\vspace{-.75em}
    \centering
    \begin{subfigure}{.33\linewidth}
      \includegraphics[width=\linewidth]{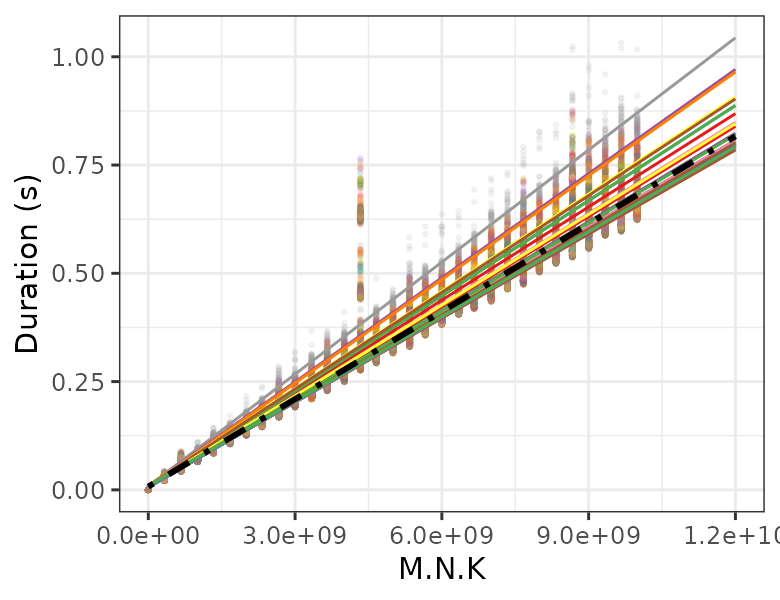}
      \caption{\texttt{dgemm} heterogeneity\label{fig:dgemm_het}}
    \end{subfigure}%
    \begin{subfigure}{.33\linewidth}
      \includegraphics[width=\linewidth]{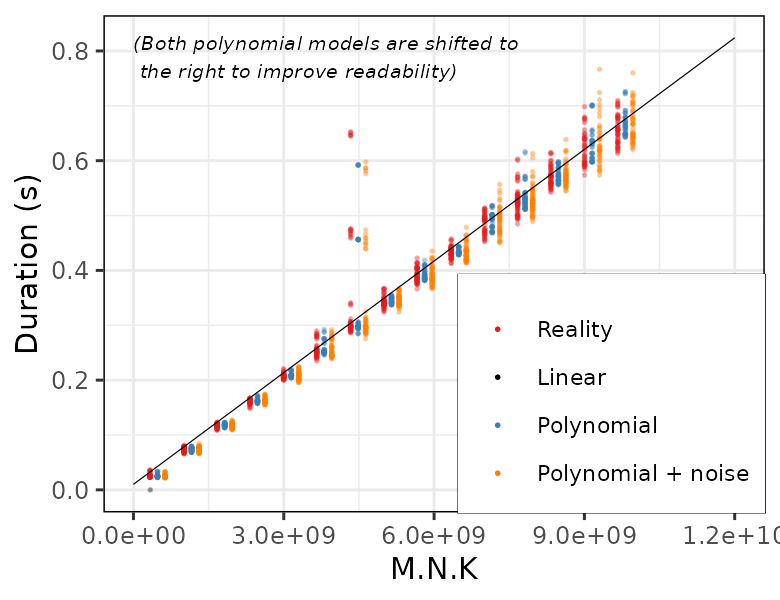}
      \caption{\texttt{dgemm} model\label{fig:dgemm_poly}}
    \end{subfigure}%
    \begin{subfigure}{.33\linewidth}
      \includegraphics[width=\linewidth]{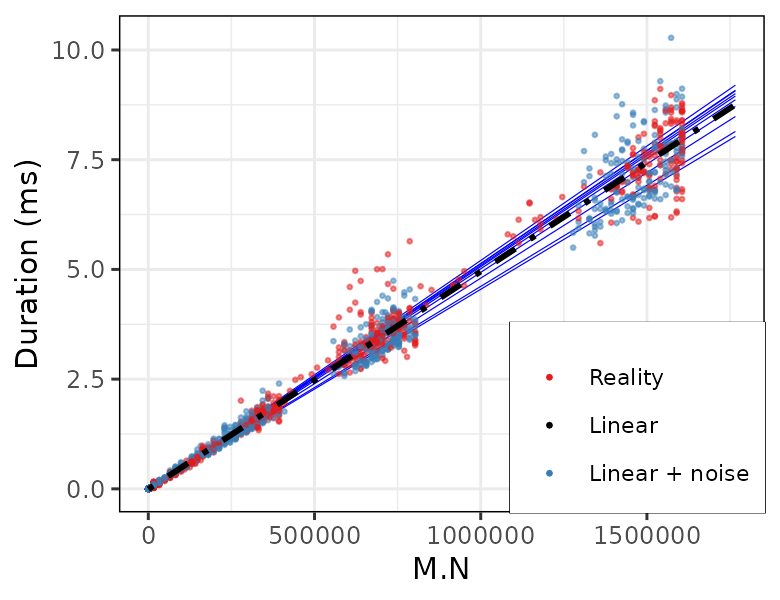}
      \caption{\texttt{HPL\_dlatcpy} model\label{fig:HPL_var}}
    \end{subfigure}
    \caption{Illustrating the realism of modeling for BLAS and HPL functions.}\vspace{-1em}
    \label{fig:blas_var}
  \end{minipage}\hfill%
\end{figure*}

\label{sec:modeling}
Most BLAS kernels have several parameters from which a straightforward
model can generally easily be identified (\eg proportional to the
product of the parameters), but refinements including the individual
contribution of each parameter as well as the \emph{spatial} and \emph{temporal}
variability of the operation are also possible. In the following, all
the simulations have been done with the following model for the \texttt{dgemm}
kernel:
\begin{equation}
  \label{eq:dgemm.complex}
  \begin{split}
    \text{For each processor $p$, } \textsf{dgemm}_{p}(M, N, K) \sim \mathcal{H}(\mu_{p}, \sigma_{p})\\
    \begin{cases}
      \mu_p &= \alpha_pMNK + \beta_pMN + \gamma_pMK + \delta_pNK + \epsilon_p\\
      \sigma_p &= \omega_pMNK + \psi_pMN + \phi_pMK + \tau_pNK + \rho_p
    \end{cases},
  \end{split}
\end{equation}
where \(\mathcal{H}(\mu, \sigma)\) denotes a half-normal random variable with
parameters \(\mu,\sigma\) accounting for the expectation and the standard
deviation. The dependency on \(p\) allows to account for platform
heterogeneity (since \(\alpha_p,\beta_p, \dots,\rho_p\) can be specific to each node),
\ie the aforementioned spatial variability. Figure\ref{fig:dgemm_het}
illustrates the importance of distinguishing between nodes: each color
and each regression line under a simple linear model corresponds to a
different cpu, whereas the black dotted line corresponds to a
regression line over all the nodes. Figure\ref{fig:dgemm_poly}
illustrates the gain brought by a fully polynomial model (blue) over a
simple linear model (black) for a given node. Indeed, for \(M.N.K \approx 4.5 \times 10^9\)
some duration are systematically higher regardless of the node. In
this particular set of experiments the corresponding combination of \(M,N,K\)
corresponds to some particular (e.g., tall and skinny) matrix geometry
which are better handled by a full polynomial model. Last, the \(\sigma_p\) parameter allows
to account for (short-term) temporal variability, \ie to model the
fact that the duration of two successive calls to \texttt{dgemm} with the same
parameters \(M, N, K\) are never identical. Modeling this variability is
important as it may propagate through the communication pattern of the
application (late sends and late receives). Figure\ref{fig:dgemm_poly}
illustrates the gain brought by modeling this variability (orange). The rationale for using a
half-normal distribution rather than a normal distribution stems from
the natural positive skewness of compute kernel duration. This model is
much more complex than the simple deterministic one used in
Figure\ref{fig:macro_simple} but, as we will explain, this complexity
is key to obtain good performance
prediction\cite{cornebize:hal-02096571}. There are four other BLAS
kernel (e.g., \texttt{daxpy}) and a few HPL kernels (often related to memory
management) but their total duration represents a negligible fraction
of the overall execution time, which 
have been modeled with a simple deterministic and homogeneous model
such as 
$\texttt{daxpy}(N) = \alpha N + \beta$ or $\texttt{HPL\_dlatcpy}(M,N) = \alpha M.N + \beta$ (see Figure\ref{fig:HPL_var}).

\subsection{Experimental Setup}
\label{sec:org124b3f5}
\label{sec:methodology.setup}
To evaluate the soundness of our approach, we compare several real
executions of HPL with simulations using the previous models.  We used
the Dahu cluster from the Grid'5000 testbed. It has 32 nodes connected
through a single switch by \SI{100}{\giga\bit\per\second} Omnipath
links. Each node has two Intel Xeon Gold 6130 CPUs with 16 cores per
CPU, and we disabled hyperthreading.  We used HPL version 2.2 compiled
with GCC version 6.3.0. We also used the libraries OpenMPI version
2.0.2 and OpenBLAS version 0.3.1.  Unless specified otherwise, HPL
executions were done using a block size of 128, a matrix of varying
size (from \num{50000} to \num{500000}), one single-threaded MPI rank
per core, a look-ahead \texttt{depth} of 1, and the \texttt{increasing-2-ring} broadcast
with the \texttt{Crout} panel factorization algorithms as this is the
combination that led to the best performance overall.
Although this machine is much smaller than top supercomputers,
faithfully simulating an HPL execution with such settings
is quite challenging.
\begin{itemize}
\item We used one rank per core to obtain a higher number
(1024) of MPI processes. This configuration is more difficult than simulating
one rank per node, as (1) it increases the amount of data
transferred through MPI, and (2) the performance is subject to
memory interference and network heterogeneity (intra-node
communications vs. inter-node communications).
\item We used a much smaller block size than what is commonly used, which
leads to a higher number of iterations and hence more complex
communication patterns.
\item We used relatively small input matrices, which reduces the makespan
and makes good predictions harder to obtain.
\end{itemize}
\subsection{A First Validation}
\label{sec:orgf96d880}
\label{sec:validation.single}
\begin{figure}[pt]
\begin{minipage}{0.8\linewidth}
    \centering
    \includegraphics[width=\linewidth]{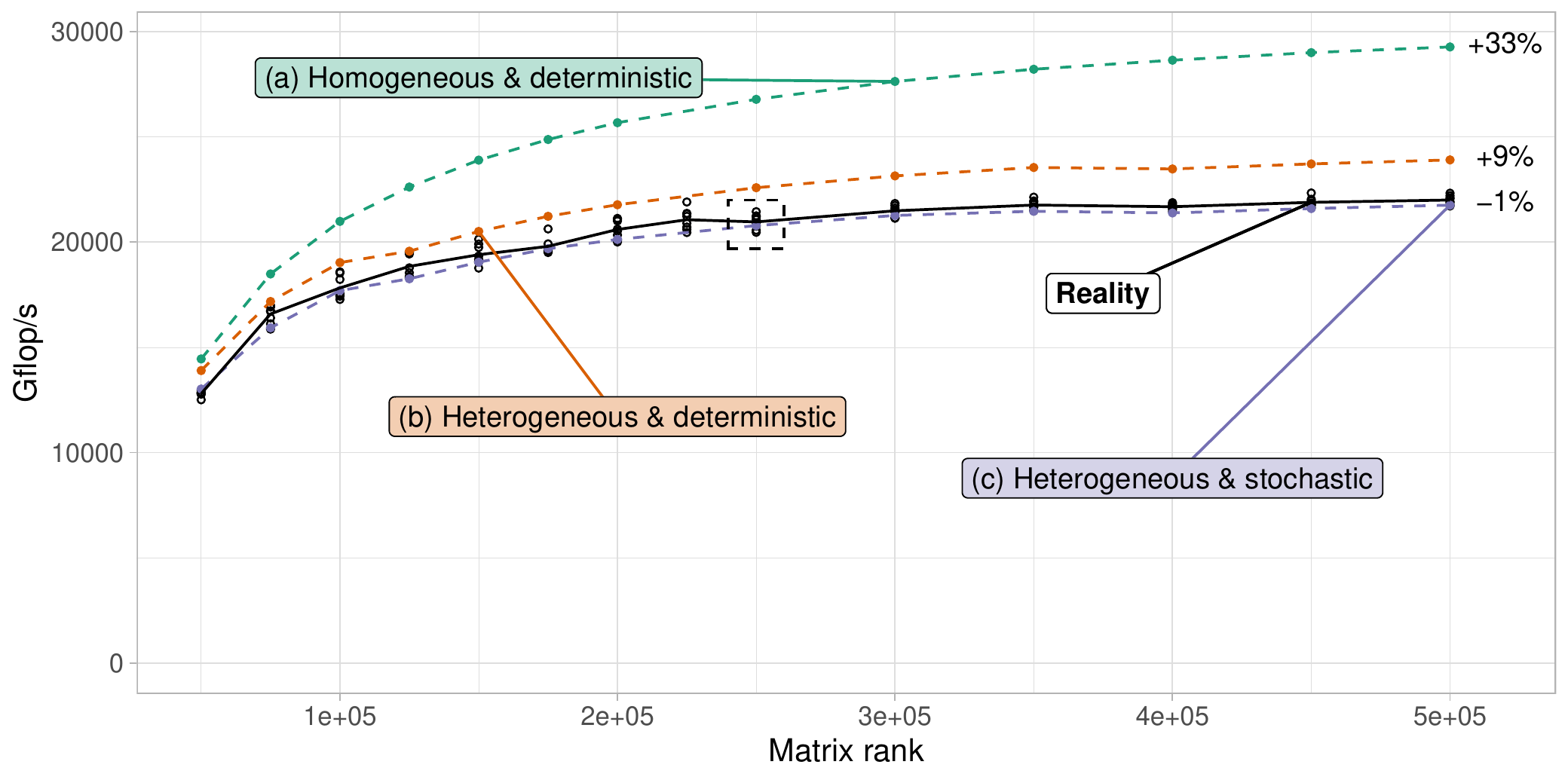}

\end{minipage}%
\begin{minipage}{0.2\linewidth}
\vspace{-1em}
\scalebox{.8}{\begin{tabular}{l|l}
\texttt{NB}        & \Num{128}          \\
\texttt{P}$\times$\texttt{Q}             & 32$\times$32        \\
\texttt{PFACT}     & Crout         \\
\texttt{RFACT}     & Right         \\
\texttt{SWAP}      & Binary-exch.  \\
\texttt{BCAST}     & 2 Ring        \\
\texttt{DEPTH}     & 1             \\
\end{tabular}}
\end{minipage}
    \caption{HPL performance: predictions (dashed lines) vs. reality (solid lines).}
    \vspace{-1em}
    \label{fig:validation_performance}
    \labspace
\end{figure}

We now present a first quantitative comparison of the prediction with
the reality on a typical scenario.
We report in Figure\ref{fig:validation_performance} the
\si{\giga\flops} rate reported by HPL when varying the matrix size \(N\).
Real executions are depicted in solid black, and the natural
variability of the overall performance is illustrated by reporting eight
runs of HPL for each matrix size. The dashed line (a), on top, is our
first attempt to simulate HPL with the naive model (homogeneous and
deterministic for both the kernels and for the network) illustrated in
Figure\ref{fig:macro_simple}. This model
overestimates HPL performance by more than \SI{30}{\percent}.
Modeling the heterogeneity of \texttt{dgemm} (\ie introducing the
dependency on \(p\) for \texttt{dgemm} as done in Eq\eqref{eq:dgemm.complex} but
without the temporal variability induced by \(\sigma_p\)) increases 
significantly the realism of the simulation as the performance is then
overestimated by only \SI{9}{\percent} (dashed line (b)).
Finally, we found that adding the temporal variability is the key ingredient to
obtain the last bit of realism.  The prediction using the
full-fledged model (dashed line (c)) is extremely 
close to reality as it slightly underestimates the performance by less than
\SI{5}{\percent} and even as little as \SI{1}{\percent} for the larger matrices.

As illustrated in Figure\ref{fig:validation_performance} and explained
in our previous work\cite{cornebize:hal-02096571}, accurate predictions require
careful modeling of both spatial and temporal variability, as they
appear to have a very strong effect on HPL performance. Somehow, this is
expected since HPL is an iterative program that synchronizes
through the broadcast of factorization panels. A single slower or late
process will eventually delay all the other ones. In the scenario
presented in Figure\ref{fig:validation_performance}, a large fraction
of the overall execution time is spent in MPI communications but
foremost in synchronizations (induced by late sends and let receives)
rather than in actual data transfers. As a consequence, careful
modeling of computations is essential, but careless modeling of the
network was enough to obtain good predictions. This article
presents an extensive (in)validation study that demonstrates the
importance of careful modeling of the whole platform.

\subsection{Experiment Time Frame}
\label{sec:org7d9b1a6}
\label{sec:methodology.timeframe}
This validation study has been carried out over several years (from 2018
to 2020). Despite our efforts to keep the
experimental setup stable for the sake of reproducibility, the platform has
evolved. The Linux kernel had a minor update, from version 4.9.0-6 to version
4.9.0-13, and the BIOS and firmware of the nodes have been upgraded. During this
time frame, the cluster has also suffered from hardware issues, like a
cooling malfunction on four of its nodes and several faulty memory modules that
had to be changed. This malfunction had an enormous impact on the
performance of HPL, which significantly complicated our validation
study but also makes it more meaningful as it has been conducted on a
particularly challenging setup.

\label{sec:validation.temperature}
\begin{figure}[pt]
\begin{minipage}{0.8\linewidth}
    \centering
    \includegraphics[width=\linewidth]{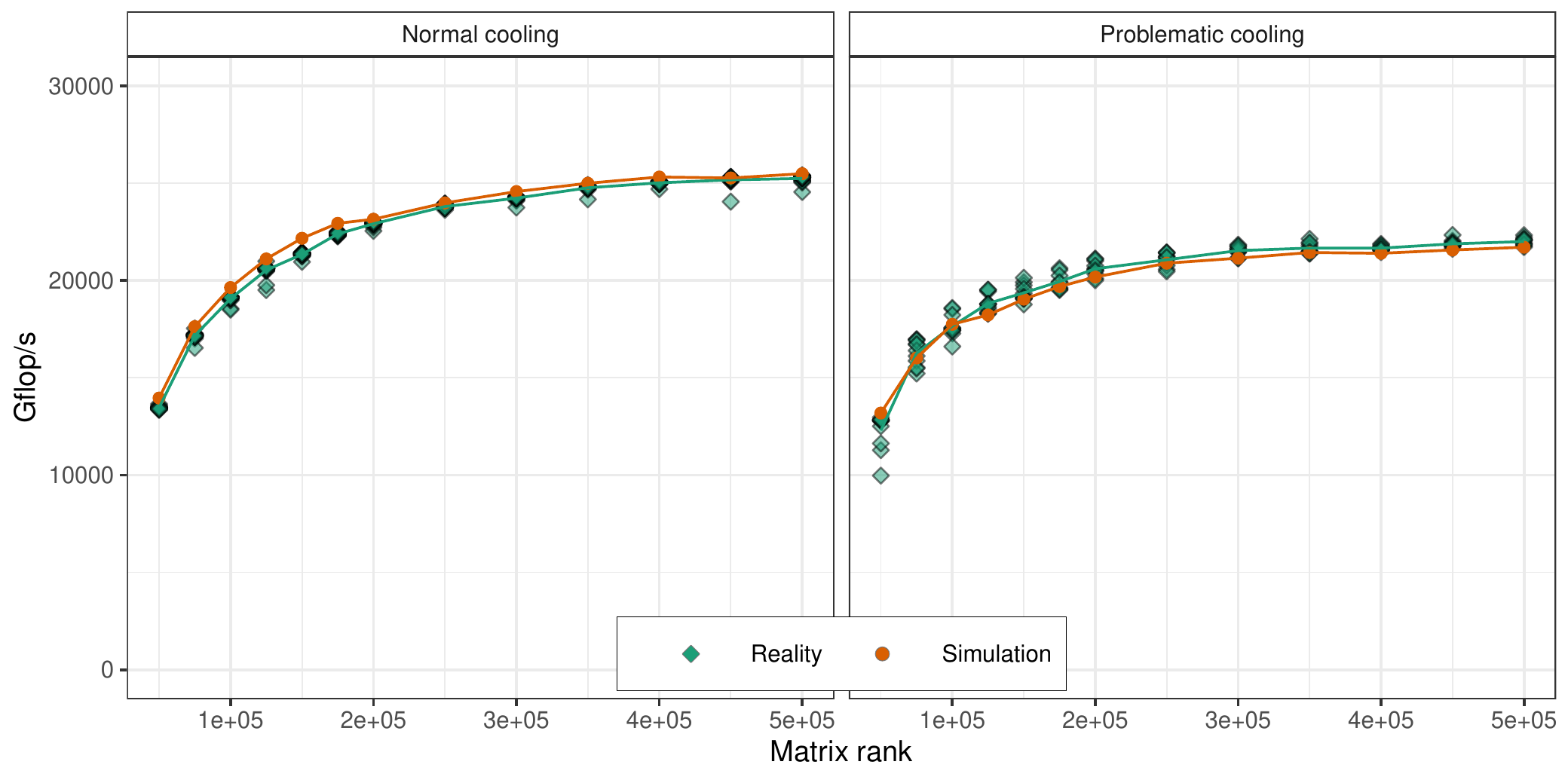}

    \vspace{-1em}
    \labspace
\end{minipage}%
\begin{minipage}{0.2\linewidth}
\vspace{-1em}
\scalebox{.8}{\begin{tabular}{l|l}
\texttt{NB}        & \Num{128}          \\
\texttt{P}$\times$\texttt{Q}             & 32$\times$32        \\
\texttt{PFACT}     & Crout         \\
\texttt{RFACT}     & Right         \\
\texttt{SWAP}      & Binary-exch.  \\
\texttt{BCAST}     & 2 Ring        \\
\texttt{DEPTH}     & 1             \\
\end{tabular}}
\end{minipage}
\caption{HPL performance: predictions vs. reality (effect of the cooling issue on the nodes dahu-\{13,14,15,16\}).}
\label{fig:validation_temperature}
\end{figure}

Our simulation approach makes it possible to predict the performance
of HPL for a new platform state by merely making a new calibration whenever a
significant change is detected.  This ability to reflect in simulation
a platform change is illustrated in
Figure\ref{fig:validation_temperature} which, similarly to
Figure\ref{fig:validation_performance} (acquired in March 2019),
showcases the influence of matrix size on the performance but at
different periods.  The left plot represents the \emph{normal}
state of the cluster (in September 2020), whereas the right plot has
been obtained (in March-April 2019) when 4 of the 32 nodes had a cooling
issue which lowered their performance by about 10\%. In all cases, we
consistently predict performance within a few percent and performing
a new \texttt{dgemm} calibration on these four nodes was all that was
needed to reflect this platform change in the simulation.

This result illustrates both the faithfulness of our simulations and a
potential use case for predictive simulations: a discrepancy between
the reality and the predictions can sometimes indicate a real issue on
the platform (similar situations have already been reported in\cite{smpi}).
\section{Comprehensive Validation Through HPL Performance Tuning}
\label{sec:orgf4c412a}
\label{sec:validation}
This section reports a few typical performance studies involving
HPL through both real experiments and simulations. The comparison of
both approaches allows us (1) to cover a broader range of parameters than
solely matrix size as done in our earlier work, (2) to evaluate how
faithful to reality our simulations are even in suboptimal
configurations, and (3) to report the main difficulties 
encountered when conducting such a study.
\subsection{Evaluating the Influence of the Geometry}
\label{sec:orgdbb9bbc}
\label{sec:validation.geometry}
\begin{figure}[t]
\begin{subfigure}{\linewidth}
    \centering
    \includegraphics[width=\linewidth]{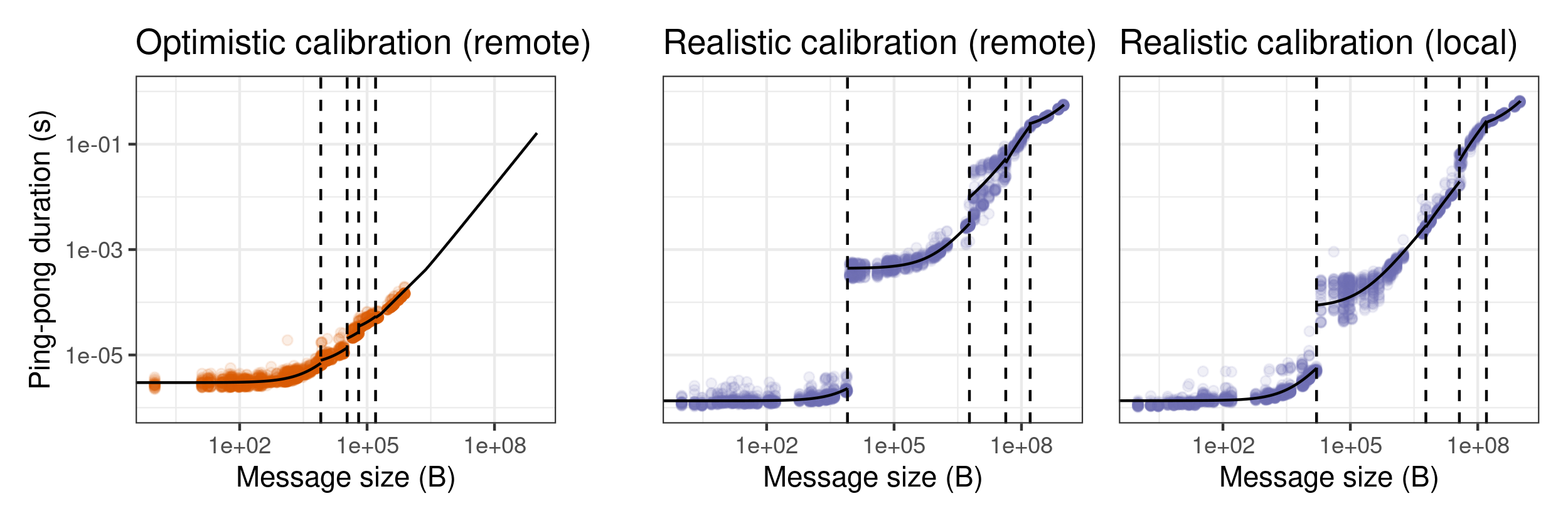}
    \caption{Illustrating the effect of the two MPI calibration
      methods.}
    \label{fig:mpi_calibration}
\end{subfigure}

\begin{subfigure}{\linewidth}
\begin{minipage}{0.8\linewidth}
    \centering
    \includegraphics[width=\linewidth]{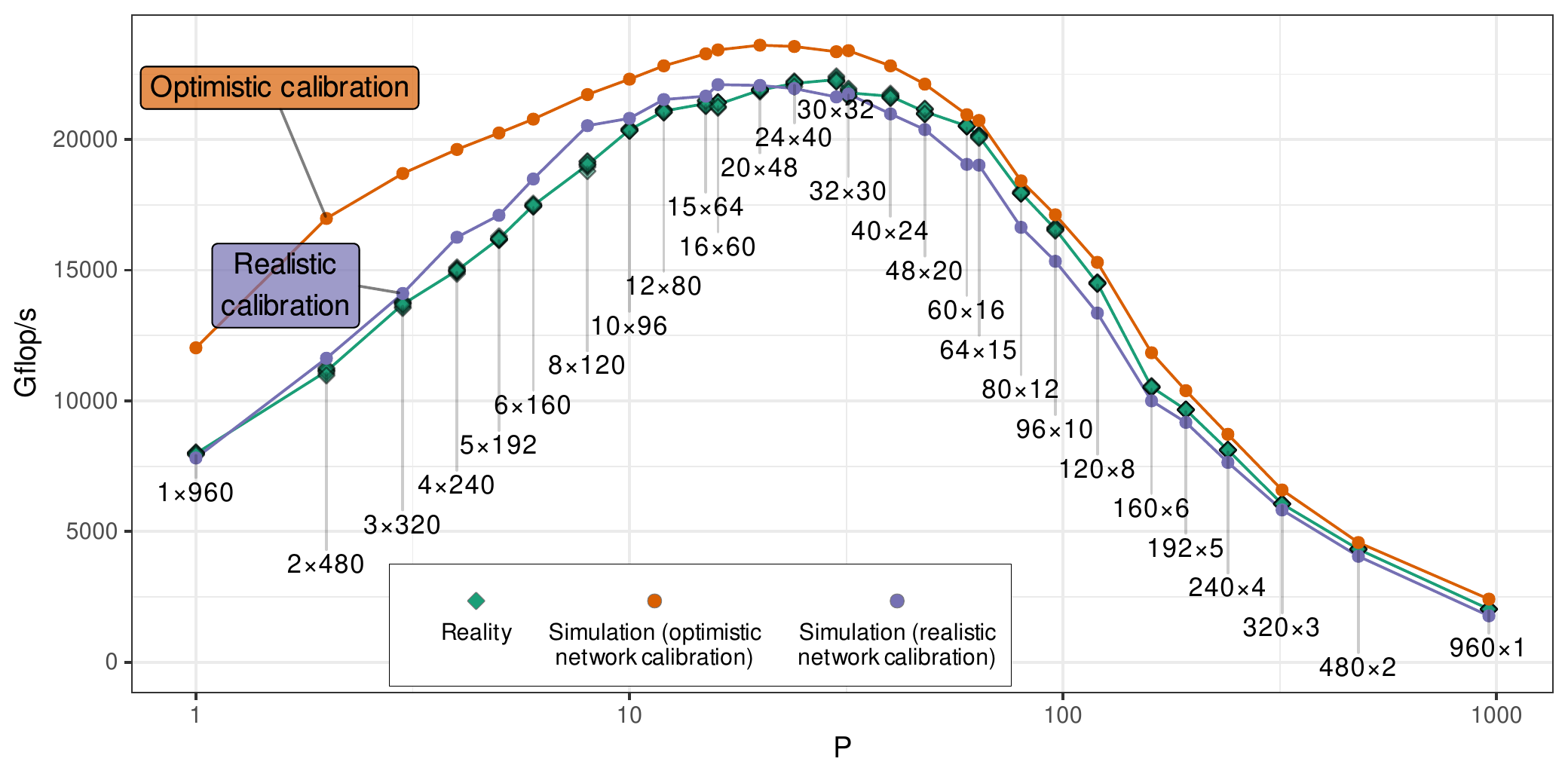}

    \vspace{-1em}
    \labspace
\end{minipage}%
\begin{minipage}{0.2\linewidth}
\vspace{-1em}
\scalebox{.8}{\begin{tabular}{l|l}
\texttt{N}         & \Num{250000}               \\
\texttt{NB}        & \Num{128}          \\
\texttt{P}$\times$\texttt{Q}             & 960        \\
\texttt{PFACT}     & Crout         \\
\texttt{RFACT}     & Right         \\
\texttt{SWAP}      & Binary-exch.  \\
\texttt{BCAST}     & 2 Ring        \\
\texttt{DEPTH}     & 1             \\
\end{tabular}}
\end{minipage}
\caption{HPL performance: predictions vs. reality (testing all the
  possible geometries for 960 MPI ranks).}
\label{fig:validation_geometry}
\end{subfigure}
\caption{The first (optimistic) network calibration gave poor
  predictions for very elongated geometries while the improved
  calibration provides perfect predictions.}
\label{fig:validation_geometry_global}
\end{figure}

Figure\ref{fig:validation_geometry} illustrates the influence on
the performance of the geometry of the virtual topology (\texttt{P} and \texttt{Q}) used
in
HPL\@.
As expected, geometries that are too distorted lead to
degraded performance.  All the HPL parameters were fixed (matrix rank
is fixed to \Num{250000} and the other parameters are the same as in
Section\ref{sec:validation.single}) except for the geometry as we
evaluate all the pairs (\texttt{P}, \texttt{Q}) such that \(\texttt{P}\times\texttt{Q}=960\).
We used only 30 nodes instead of 32 to cover a larger number of
geometries, as \Num{960} has more divisors than \Num{1024}.

As in all our previous studies, we report both the predicted
performance and the one measured in reality. Like the
comparisons presented in the previous section, the simulation was done
with the \texttt{dgemm} model from Eq\eqref{eq:dgemm.complex} (stochastic, heterogeneous, and
polynomial) and the simplest linear models for the other kernels.  In our
first simulation attempt that relied on a relatively simple network
model (deterministic yet piecewise-linear to account
for protocol switch) depicted on the leftmost plot of
Figure\ref{fig:mpi_calibration}, we obtained the unsatisfying orange
line on top of Figure\ref{fig:validation_geometry} for the prediction.
The simulations with the smallest value of \texttt{P} 
had relatively large prediction errors, with a systematic
over-estimation that reaches up to +50\% for the \(1\times960\) and \(2\times480\)
geometries.  A qualitative comparison of the execution traces obtained
in reality and simulation showed that the broadcast phases' duration
was greatly underestimated in simulation. We found
out that with such elongated geometries, the message size is
significantly larger than what we had used in our calibration, and
the performance surprisingly and significantly drops for such
size (compare with the rightmost plots of Figure\ref{fig:mpi_calibration} for messages larger than
\SI{160}{\mega\byte}). This performance drop is
explained by poor optimization of the DMA locking mechanism in the
Infiniband network layer\cite{denis:inria-00586015}. A similar
performance drop also happens for intra-node communications that
poorly manage the caches above a given size. Furthermore, the
communication patterns generated by HPL during the ring broadcast are
significantly impacted by the busy waiting of HPL that intensively
calls \texttt{MPI\_Probe} and \texttt{dgemm} on small sub-matrices. Our initial procedure
for calibrating the network did not capture this phenomenon since we
did not inject any additional CPU load.

We addressed this problem by improving our network calibration
procedure: (1) we use a distinct model for local and remote
calibrations, (2) we sample the message sizes in a larger interval (up to
\SI{2}{\giga\byte} instead of only \SI{1}{\mega\byte}), and
(3) we add calls to \texttt{dgemm} and \texttt{MPI\_Iprobe} between
each call to \texttt{MPI\_Send} and \texttt{MPI\_Recv}. The goal was
to make the calibration environment more similar to what happens in
HPL\@.
The resulting network model is illustrated in the rightmost plots of
Figure\ref{fig:mpi_calibration}. This more realistic network model
solved every previous misprediction and allows us to produce very
faithful simulations (purple line on
Figure\ref{fig:validation_geometry}), which are now a few percent of
the reality regardless of the geometry. This figure also illustrates
the influence of the geometry on overall performance since there is
almost a factor of ten between the worst configuration (\(960\times1\)) and the
best one (\(30\times32\)). Although it is not surprising to see that the
geometries which are as square as possible lead to better
performance as they minimize the overall amount of data movements, it
is interesting to observe the asymmetric role of \texttt{P} and \texttt{Q} in the
overall performance (smaller values for \texttt{P} lead to better performance)
and which can be explained by the structure of the collective
operations but requires a close look at the code.
\subsection{Optimizing the HPL Configuration Through a Factorial Experiment}
\label{sec:org6b0b53e}
\label{sec:validation.factorial}
\begin{figure}[pt]
    \centering
    \includegraphics[width=\linewidth]{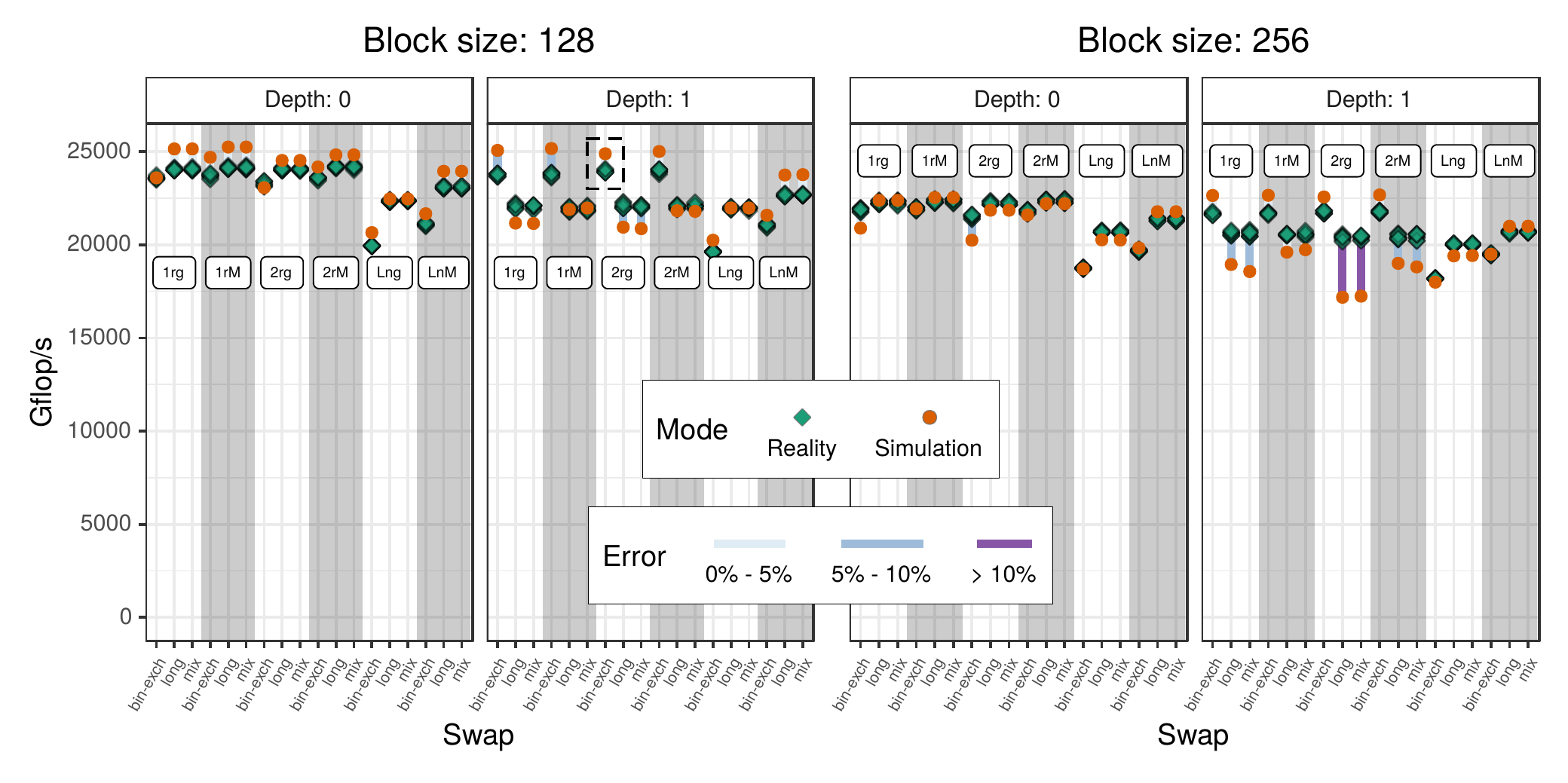}

    \caption{Influence of HPL configuration on the performance (factorial experiment).}\vspace{-1em}
    \label{fig:validation_factorial}
    \labspace
\end{figure}

Although geometry is among the most important parameter to tune, six
other parameters control the behavior of HPL. In
Figure\ref{fig:validation_factorial}, we compare the performance
reported by HPL when fixing the matrix rank to \Num{250000} and
varying the following parameters: block size (128 or 256), depth (0 or
1), broadcast (the six available algorithms), swap (the three
available algorithms). The geometry was fixed to
\(\text{P}\times\text{Q}=32\times32=1024\) as it is optimal (the simpler
calibration procedure and the network model depicted on the leftmost
plot of Figure\ref{fig:mpi_calibration} were thus used). The
parameters \texttt{pfact} and \texttt{rfact} (panel factorization)
were respectively fixed to \texttt{Crout} and \texttt{Right}, as they
had nearly no influence on HPL performance in our early
experiments.

Figure\ref{fig:validation_factorial} depicts the 72 parameter combinations we
tested. Parameters have been reorganized based on their influence on performance
to improve readability. The boxed configuration corresponds to the one boxed in
Figure\ref{fig:validation_performance}. These parameters account for up to
\SI{30}{\percent} of variability in the performance, which is less
important than the geometry but is still quite significant. For 61 of
them, the prediction error is lower than \SI{5}{\percent}. Only two
combinations have shown a large error of approximately
\SI{15}{\percent}, obtained with a block size of 256, a depth of 1,
the \texttt{2-ring} broadcast algorithm, and either the \texttt{long} or the \texttt{mix} swap
algorithm. This demonstrates the soundness of our approach, as our
predictions are reasonably accurate most of the time.  This experiment
confirms that, although the prediction of HPL performance for a given
parameter combination has a systematic bias, the error remains within
a few percent most of the time. Therefore, this surrogate is good
enough for parameter tuning and should be considered when preparing a
large-scale run.

While testing all the parameter combinations is the safest method to
discover the combination that provides the highest performance, its
cost can be prohibitive due to the high number of
combinations. An alternative often used in practice is to explore only
a small subset of the parameter space and to analyze
variance (ANOVA) to identify the parameters with the more substantial effect
on performance and then select the appropriate combination.  We
applied this procedure on samples of both datasets (the one obtained from real
runs and the one obtained in simulation). In both cases, the two
parameters with the highest effect were the block size \texttt{NB} and the
\texttt{depth}, as shown in Figure\ref{fig:validation_factorial}, followed by
\texttt{bcast} and \texttt{swap}. The best combinations selected in both cases were also
identical, demonstrating once again the faithfulness of our simulation
approach and how it can be used to reduce the experimental cost of
parameter tuning.
\subsection{Conclusion}
\label{sec:orga1b1d33}
Accurately predicting the performance of an application is not a
trivial task. Discrepancies between reality and simulation can be
multiple: the platform may have changed (\eg the cooling issue that
affected four nodes in Section\ref{sec:methodology.timeframe}), the
model could be inaccurate (\eg the homogeneous and deterministic
\texttt{dgemm} model is too simple as in\cite{cornebize:hal-02096571})
or not correctly calibrated (\eg the calibration procedure does not
cover the appropriate parameter space, or the experimental conditions
are too different as in Section\ref{sec:validation.geometry}). As
expected in any serious investigation of model validity, our validation
study is not a mere collection of positive cases. Instead, it is the
result of a thorough (we extensively covered the HPL parameter space)
attempt to invalidate our model as well as explanations on how we did so. By
meticulously overcoming each of these issues, we have demonstrated the
ability of our approach to produce very faithful predictions of HPL
performance on a given platform.
\section{Sensibility Analysis in What-if Scenarios}
\label{sec:org7906a2d}
\label{sec:whatif}
We have shown that many typical HPL case studies could be conducted in
simulation. However, their conclusions (optimal geometry and
parameters) are specific to the cluster we used and they require a
precise model of several aspects of the target cluster, which may not
be possible at early experimental stages. In particular, only a few
cluster nodes may be available at first and the whole cluster model
should then be constructed from a limited set of observations and
carefully extrapolated. This section shows how typical \emph{what-if}
simulation studies should be conducted given such
uncertainty. Section\ref{sec:whatif.model} presents a generative model
of node performance that can easily be fit from daily measurements and
used to produce a similar platform. This model is used to quantify
the importance on overall performance of temporal variability of the
\texttt{dgemm} kernel in Section\ref{sec:whatif.temporal_variability} and of
spatial variability of nodes in
Section\ref{sec:whatif.spatial_variability}. In particular, we show
how to study the efficiency of a simple \emph{slow node eviction}
strategy. Finally, we study in Section\ref{sec:whatif.topology} the
influence of the physical network topology on overall
performance. Most of these studies are particularly difficult to
conduct through real experiments because of the difficulty to finely
control the platform.
\subsection{A Generative Model of Node Performance}
\label{sec:org0ecb574}
\label{sec:whatif.model}
As we have seen in Section\ref{sec:modeling}, the performance of nodes
exhibits several kinds of variability: i) a spatial
variability (between nodes) ii) a ``short-term'' temporal variability
(the one experienced within an HPL run) but also iii) a ``long term''
temporal variability (from a day to another). As illustrated in
Section\ref{sec:validation.single}, accounting for the first two kinds
of variability is essential but during our investigation of the simulation
validity, which spanned over several months, we also had to deal with
the fact that the node performance from a day to another could
significantly vary, thereby making our comparisons between a real
experiment and the simulation driven by model obtained with past
measurements sometimes irrelevant.

This section explains how all sources of variability can be
accounted for in a single unified model. From our observations, we
assume that on a given node \(p\) and a given day \(d\), the duration of
the \texttt{dgemm} kernel can be modeled as follows:
\begin{equation}
  \label{eq:dgemm.basic}
  \forall M,N,K, \textsf{dgemm}_{p,d}(M, N, K) \sim \mathcal{H}(\alpha_{p,d}MNK + \beta_{p,d}, \,\, \gamma_{p,d}MNK)\\
\end{equation}
Compared to the model\eqref{eq:dgemm.complex}, this model
includes the daily variability but drops the complexity of a
full-fledged polynomial. Such complexity may be important whenever
trying to model a particular platform. However, when performing sensibility
analysis, a simpler model is preferred, especially as not all terms of
the polynomial may be statistically significant. In this model, the
short-term temporal variability stems from the \(\gamma_{p,d}\) term while the
average performance of the node stems from the \(\alpha_{p,d}\) and \(\beta_{p,d}\)
terms, which we gather in a single 3-dimensional vector
\begin{equation}
  \mu_{p,d}=(\alpha_{p,d},\beta_{p,d},\gamma_{p,d}).
\end{equation}
Now, since every machine is unique it is natural to assume that for
each machine:
\begin{equation}
  \label{eq:dgemm.temporal}
  \forall d, \mu_{p,d} \sim \mathcal{N}(\mu_{p},\Sigma_T)
\end{equation}
In this model, \(\mu_{p}\) accounts for the average performance of the
machine \(p\), while \(\Sigma_T\) accounts for its day-to-day variability. From
our observation we had no particular reason to assume that this
variability was different from a machine to another, hence, \(\Sigma_T\) is not
indexed by \(p\) but global to all machines. However, the parameters
\(\alpha_{p,d},\beta_{p,d},\gamma_{p,d}\) are generally correlated to each others,
hence \(\Sigma_T\) is full covariance matrix to account for interactions.
The choice of a Normal distribution is natural since it is the simpler
distribution that accounts for a specific mean and variance, but
we will discuss its relevance later in this section.

\begin{figure}[pt]
    \centering
    \includegraphics[scale=.911]{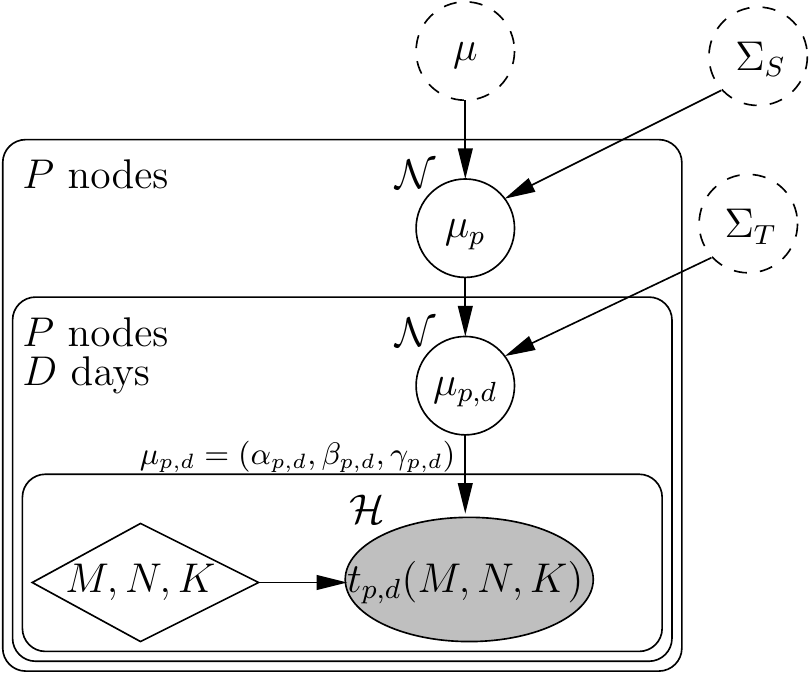}

    \caption{Generative model of kernel duration accounting for the
      spatial ($\Sigma_S$), long-term ($\Sigma_T$) and short-term variability ($\gamma_{p,d}$).}\vspace{-1em}
    \label{fig:generative}
    \labspace
\end{figure}

Finally, we need to account for the spatial variability, which we
propose to model as follows:
\begin{equation}
  \label{eq:dgemm.spatial}
  \forall p, \mu_{p} \sim \mathcal{N}(\mu,\Sigma_S)
\end{equation}
Again, in such a model \(\mu\) accounts for the machines' average
performance while \(\Sigma_S\) accounts for the (weak) heterogeneity. This
hierarchical model is depicted in Figure\ref{fig:generative}.
The shaded node represents observed variables and diamond node represents
deterministic variables, while non-shaded nodes represent latent variables. The
solid node is the variable which is estimated when conducting (in)validation
studies while the dashed ones are useful when conducting sensibility analysis
and extrapolating to an hypothetical cluster.

\begin{figure}[pt]
    \centering
    \begin{subfigure}{\textwidth}
        \centering
        \raisebox{2em}{\rotatebox{90}{\fbox{\vphantom{y}Real data}}}~%
        \includegraphics[width=.48\linewidth]{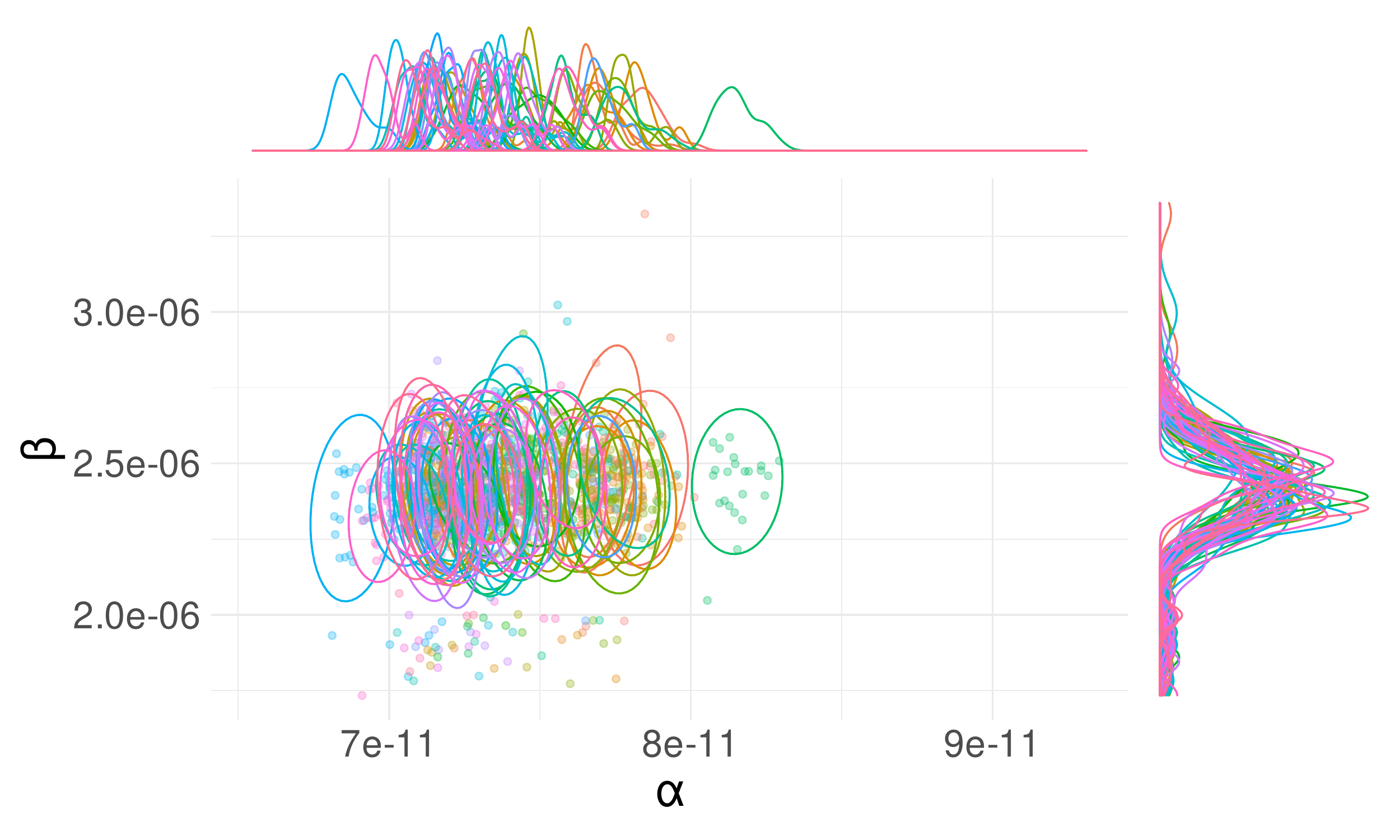}%
        \includegraphics[width=.48\linewidth]{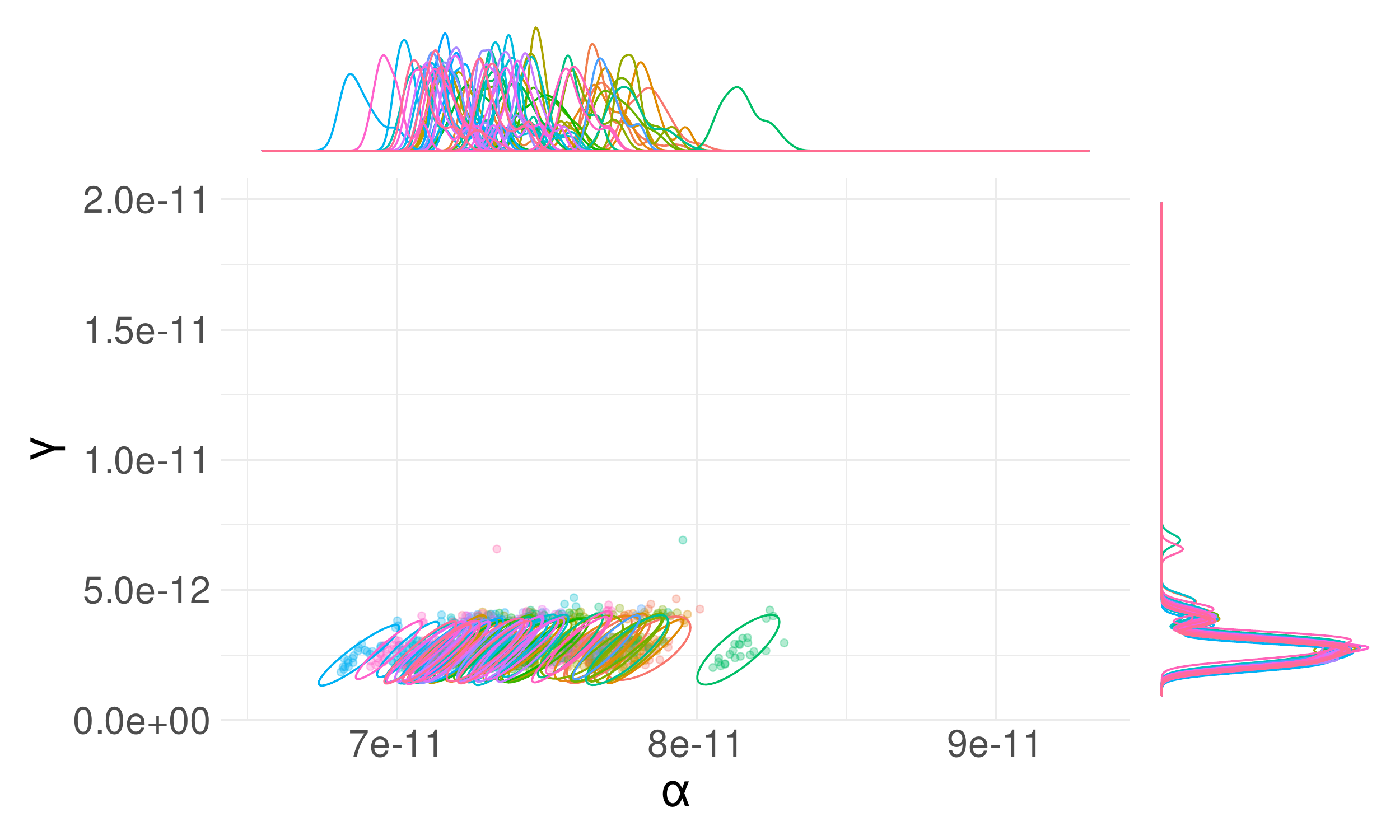}
        \caption{Distribution of $\alpha$, $\beta$, and $\gamma$ (observations on $2\times32$ CPUs from November 2019 to February 2020).}
        \label{fig:whatif_calibration}
    \end{subfigure}

    \begin{subfigure}{\textwidth}
        \centering
        \raisebox{1em}{\rotatebox{90}{\fbox{Synthetic data}}}~%
        \includegraphics[width=.48\linewidth]{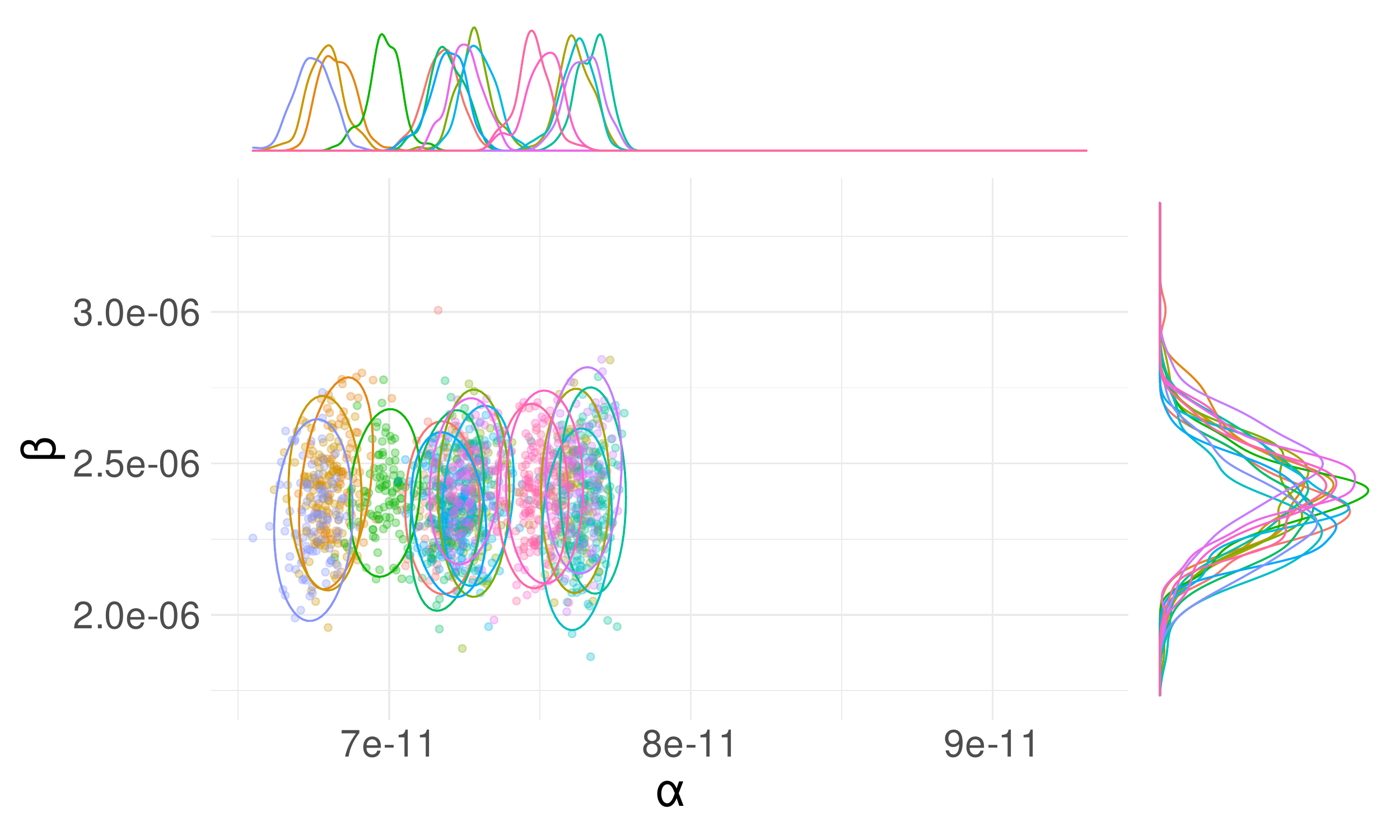}%
        \includegraphics[width=.48\linewidth]{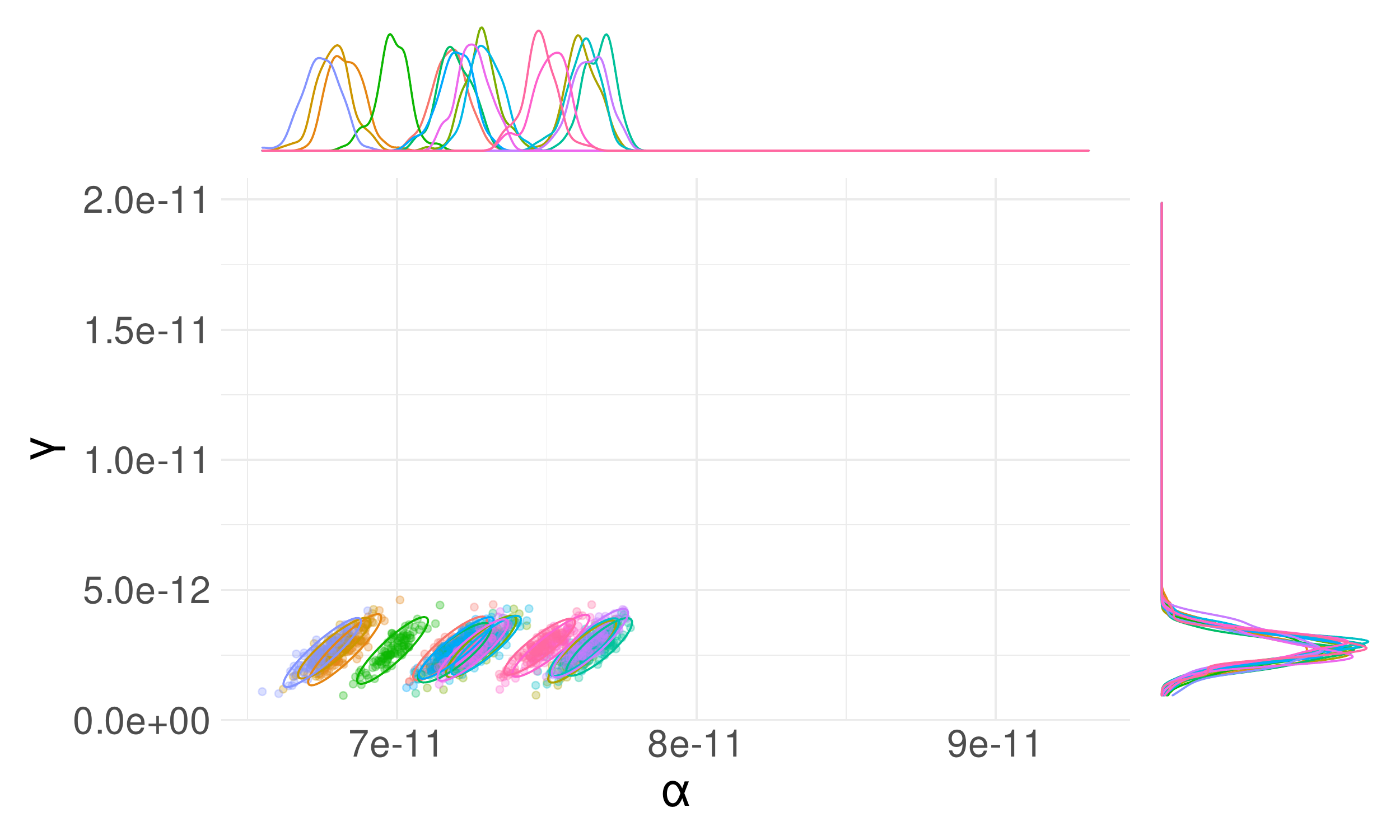}
        \caption{Distribution of $\alpha$, $\beta$, and $\gamma$ (synthetic data for 16 CPUs).}
        \label{fig:whatif_model}
    \end{subfigure}%
    \caption{Distribution of the regression parameters for around 20
      \texttt{dgemm} calibrations made on each of the 32 nodes. Each
      color/ellipse corresponds to a different CPU.}
    \label{fig:whatif}
    \labspace
    \centering
    \begin{subfigure}{\textwidth}
        \centering
        \raisebox{2em}{\rotatebox{90}{\fbox{\vphantom{y}Real data}}}~%
        \includegraphics[width=.48\linewidth]{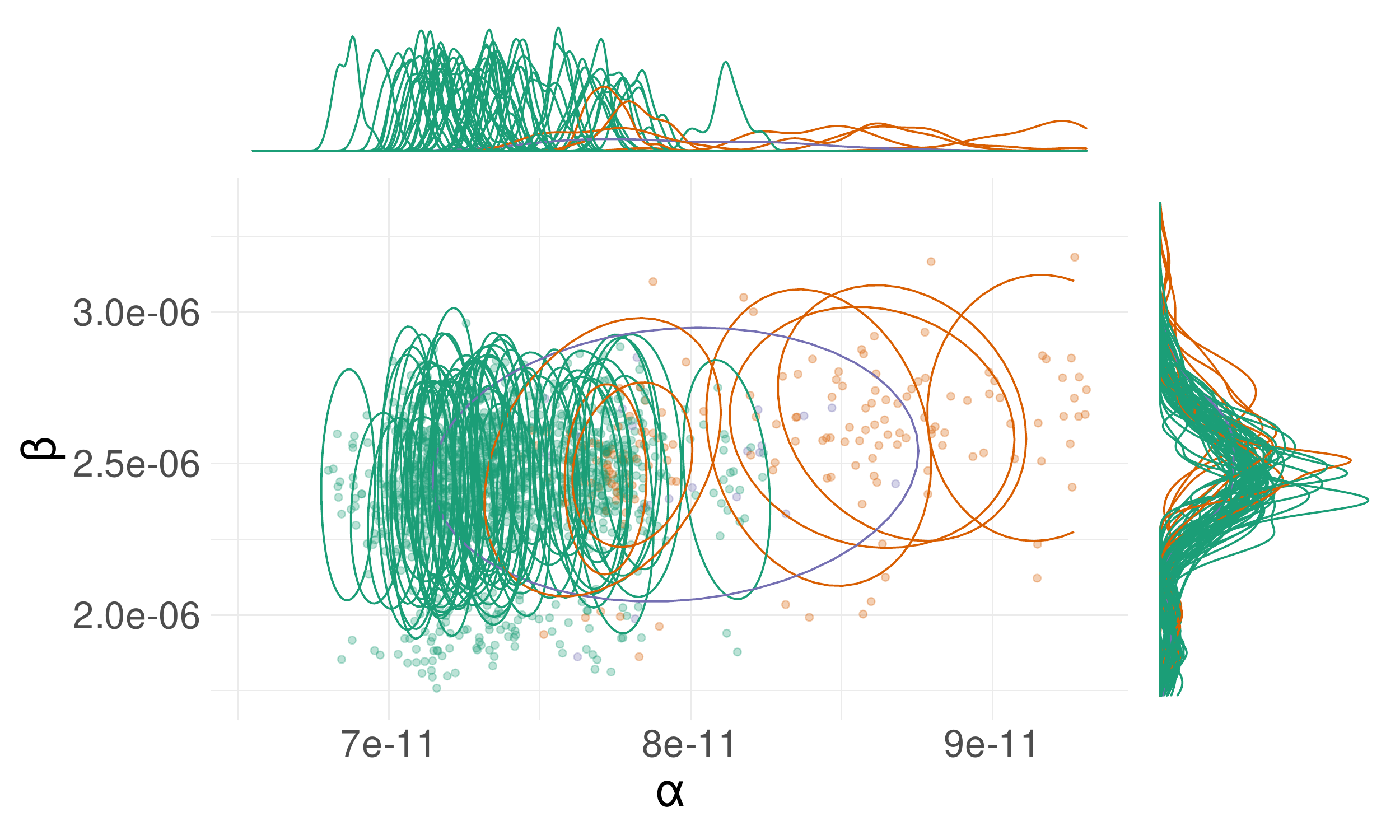}%
        \includegraphics[width=.48\linewidth]{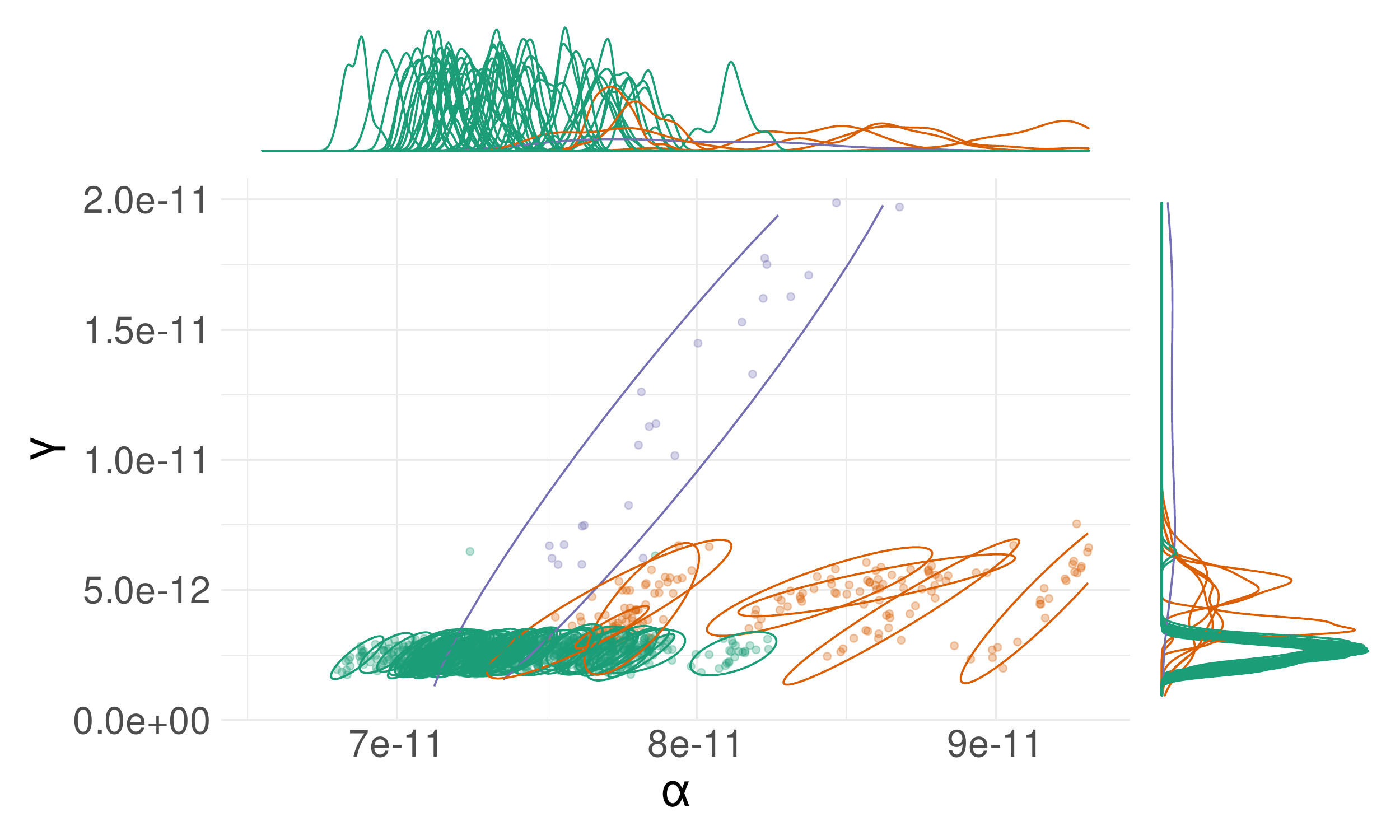}
        \caption{Distribution of $\alpha$, $\beta$, and $\gamma$ (observations on $2\times32$ CPUs from October to November 2019).}
        \label{fig:whatif_slow_calibration}
    \end{subfigure}

    \begin{subfigure}{\textwidth}
        \centering
        \raisebox{1em}{\rotatebox{90}{\fbox{Synthetic data}}}~%
        \includegraphics[width=.48\linewidth]{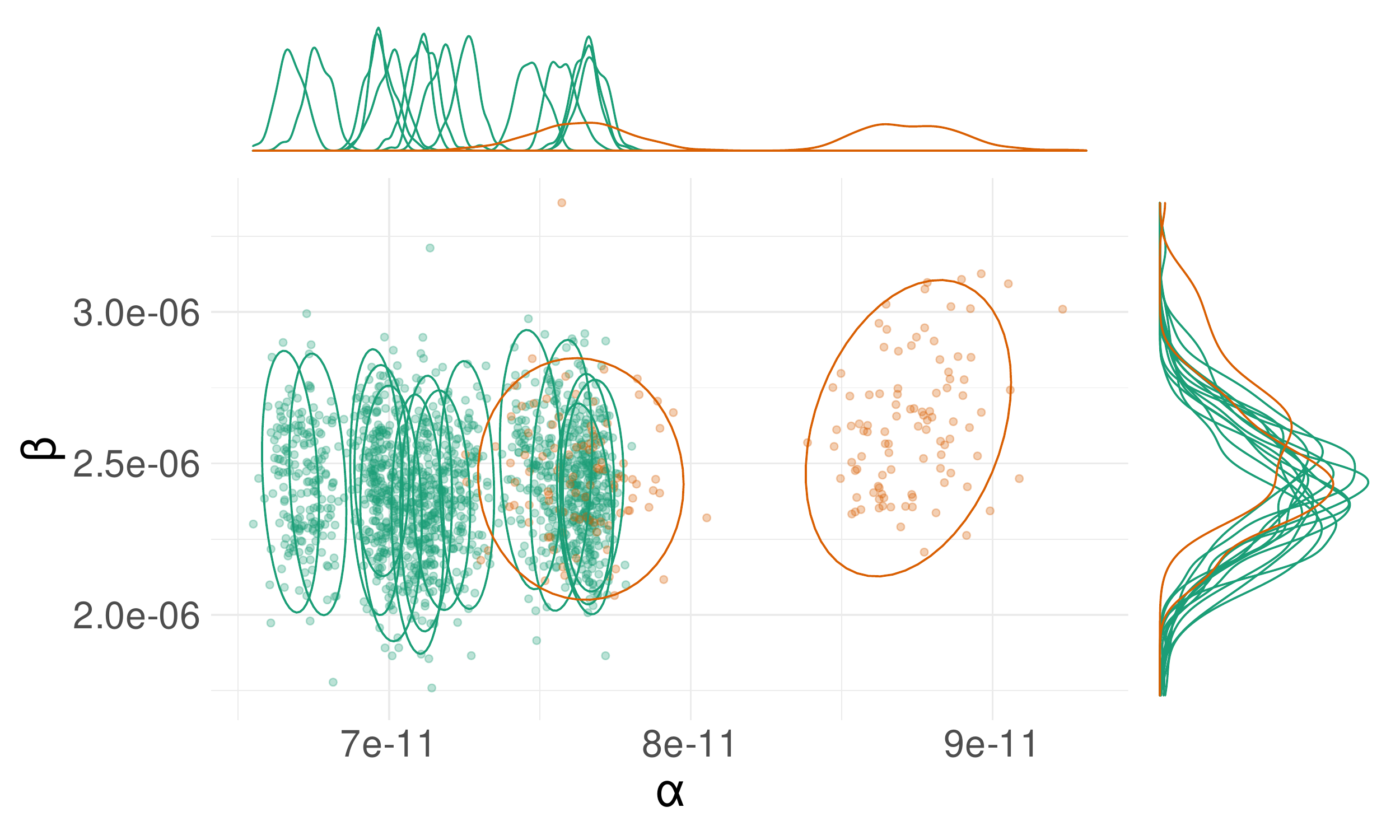}%
        \includegraphics[width=.48\linewidth]{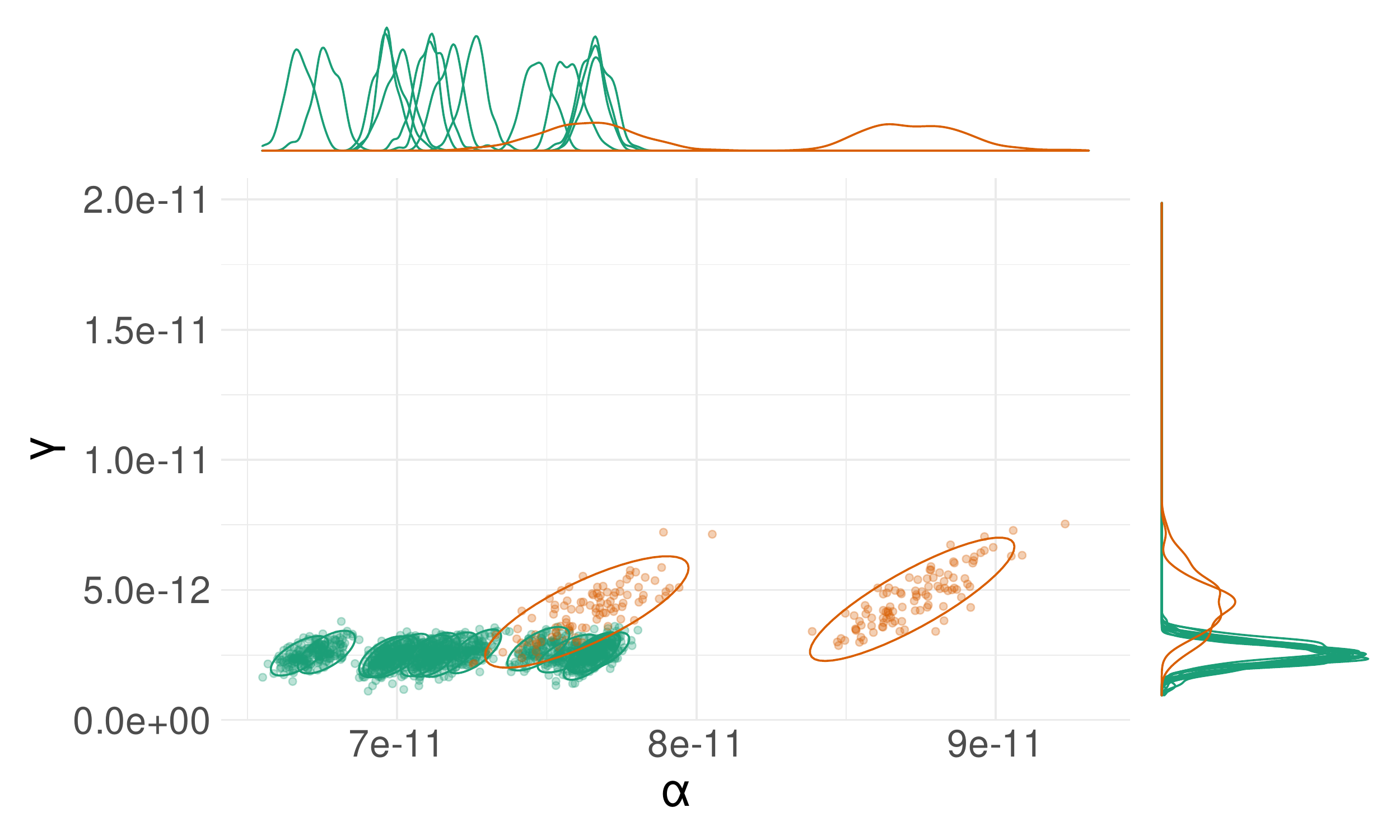}
        \caption{Distribution of $\alpha$, $\beta$, and $\gamma$ (synthetic data for 16 CPUs).}
        \label{fig:whatif_slow_model}
    \end{subfigure}%
    \caption{Same as Figure~\ref{fig:whatif}, except that 4 of
      these nodes had a cooling problem, leading to longer and more
      variable durations.}
    \label{fig:whatif_slow}
    \labspace
\end{figure}

The relevance of model\eqref{eq:dgemm.basic} has already been
illustrated in Section\ref{sec:validation.single},
\ref{sec:methodology.timeframe}, and\ref{sec:validation} but the
relevance of models\eqref{eq:dgemm.temporal}
and\eqref{eq:dgemm.spatial} requires some
attention. Figure\ref{fig:whatif_calibration} represent the empirical distribution of
\(\mu_{p,d} = (\alpha_{p,d},\beta_{p,d},\gamma_{p,d})\) (the result of the linear
regression) for the 32 nodes of the Dahu cluster on 40 different days
from November 2019 to February 2020. The distribution for each node
appears approximately normal and passed a Shapiro-Wilk normality
test. Although the distribution of the \(\beta_{p,d}\) appears slightly skewed
toward larger values and one of the nodes (the one with the larger
\(\alpha_{p,d}\)) stands out, there is no good reason for using a more complex
distribution than a Gaussian one. Although the correlation between
\(\alpha\), \(\beta\), and \(\gamma\) is very weak, it appeared to be statistically
significant (most ellipses are slightly tilted toward North-East),
hence a full variance matrix is needed (at least for \(\Sigma_T\)).

\begin{table}[b]
\caption{\(R^2\) values obtained with the linear regressions for \texttt{dgemm} durations, with both the linear and polynomial models.}
\vspace{-1em}
\label{tab:rsquared}
\center
\begin{tabular}{l|cc}
 & Linear & Polynomial\\
\hline
per host and day (\(\mu_{p,d}\)) & $[0.9842, 0.9994]$ & $[0.9958, 0.9998]$\\
per host (\(\mu_{p}\))           & $[0.9960, 0.9971]$ & $[0.9994, 0.9997]$\\
global (\(\mu\))                 & $0.9963$           & $0.9996$\\
\end{tabular}
\end{table}


Although fitting a full polynomial model for each day and each
processor provides the best predictions it is important to understand
that even cruder models are quite good. Table\ref{tab:rsquared}
provides indication about the quality of the prediction (the
coefficient of determination \(R^2\), which is comprised between 0 and 1,
with 1 indicating a systematically perfect prediction) for the 32 Dahu
nodes over the November 2019 to February 2020 period. Regardless of
the model complexity (linear/polynomial) and of the granularity of the
regression (global, per host, per host and day) the model quality is
excellent and above \(0.99\), so it could appear that there is no need
for a complex model. Yet, we have seen in
Section\ref{sec:validation.single} that a simple and global linear
model of \texttt{dgemm} which is yet an excellent microscopic model
(\(R^2=0.9963\)) fails to provide good macroscopic prediction of
HPL. The spatial variability is essential, just like the short-term
and daily variability, which is why modeling the variability of \(\mu_{p,d}\)
is key. Last, we recall that if a full polynomial model (with 10
parameters as in\eqref{eq:dgemm.complex} instead of 3 as
in\eqref{eq:dgemm.basic}) is particularly suited to predict the
performance of a specific machine and day, it is inadequate to extrapolate
performance in general as the covariance matrices \(\Sigma_T\) and \(\Sigma_S\)
would then have \(10\times9/2=45\) parameters instead of \(3\). This is why
in the following, we opt for a simpler linear model.

It is easy to estimate \(\mu_{p}\) and \(\Sigma_T\) by averaging over the
\(\mu_{p,d}\) of each node, and then to estimate \(\mu\) and \(\Sigma_S\) by averaging
over all the nodes. This moment-matching method is simple and provides
very good estimates for \(\mu\), \(\Sigma_T\), and \(\Sigma_S\) because we have enough
measurements at our disposal and because it is particularly suited to
the Gaussian modeling assumption. Should more complex models (e.g., a
mixture to account for ``outlier'' nodes or a SkewNormal distribution to
account for the distribution's skewness) be used, a general
Bayesian sampling framework like STAN\cite{stan} would be more
adapted. Such frameworks allow to easily specify hierarchical generative models
like the one presented in Figure\ref{fig:generative} and to draw
samples from the posterior distribution of \(\mu\), \(\Sigma_T\), and \(\Sigma_S\),
which can be used to generate realistic \(\mu_{p,d}\) values for a new
hypothetical cluster easily.

Such a process is depicted in Figure\ref{fig:whatif_model} where
hypothetical regression parameters for 16 nodes have been
generated. Comparing such synthetic data with the original samples from
Figures\ref{fig:whatif_calibration} allows us to evaluate the model's
potential weaknesses. Although the orders of magnitude of all
parameters and the ellipses are excellent, a few subtle differences
are visible. First, the variability of \(\alpha_{p}\) seems a bit
overestimated (the spread along the x-axis is larger). This can be
explained by the fact that one of the nodes seemed to be significantly
slower (with much larger \(\alpha_{p}\)), which artificially increased the
spatial variability. Second, as expected from a Gaussian model, the
distributions of the \(\beta_{p,d}\) are symmetrical whereas there was a
slight negative skew in the original samples but this should be of little
significance for our study. The distributions of the \(\gamma_{p,d}\) however
are particularly realistic.

We also illustrate the generality of this model with the data from
Figure\ref{fig:whatif_slow_calibration}. These measurements were
obtained from October to November 2019 where the cluster was less
stable and where some nodes particularly misbehaved. Three nodes (in
orange, hence a total of 6 CPUs) are distinguished from the 28 others
(in green) and have lower performance (higher values for \(\alpha\), \(\beta\), and \(\gamma\))
and one node (in blue) is particularly unstable. Although this last
node may be considered too abnormal to represent
anything, it would be reasonable to assume that a larger cluster would present
at least the two kinds of behaviors (green for stable nodes, and orange for
slower nodes). The higher layer of the model in
Figure\ref{fig:generative} should then be replaced by a mixture of
normal distributions (whose weights would then be sampled from a
Dirichlet distribution). Again, hypothetical regression parameters for
16 CPUs have been generated with such a process on
Figure\ref{fig:whatif_slow_model} are very similar, although
different, to the original measurements.

Overall this model is therefore of excellent quality and can be used
to generate large configurations very easily and evaluate the
influence of different kinds of variability on the performance of HPL.

\subsection{Influence of \texttt{dgemm}'s Temporal Variability}
\label{sec:org2cb3bd8}
\label{sec:whatif.temporal_variability}
In Section\ref{sec:validation.single}, we could highlight the
importance of accounting for temporal variability of the \texttt{dgemm} kernel
to obtain faithful HPL predictions. To the best of our knowledge, HPL
developers and experts are often aware of this influence (or at least
suspect it). However, they have never fully quantified it since
designing and performing real experiments to evaluate this would be
quite difficult. Although increasing this variability wouldn't be too
hard, reducing it would be particularly complicated. This can however
easily be done through simulation using the hierarchical model of the
previous section. In our experiments, the order of magnitude of
the temporal variability with respect to actual performance (i.e.,
the ratio between \(\gamma_{p,d}\) and \(\alpha_{p,d}\) in
Equation\eqref{eq:dgemm.basic}) was around 3\%. This may be a
``normal'' value or could be considered too high and possibly
improved by better controlling thread mapping or Operating System
noise. Such a task can be quite tedious and knowing how much performance
gain can be expected beforehand is thus quite useful. In this section,
we study the influence of this variability by generating 10 cluster
scenarios using the previous model (as in Figure\ref{fig:whatif}),
comprising 1,024 nodes each, but by
constraining \(\gamma_{p,d}\) to be equal to \(\gamma.\alpha_{p,d}\) with \(\gamma\in[0,0.1]\), which
represents the coefficient of variation of the \texttt{dgemm} kernel. We
evaluate the performance of HPL with one multi-threaded MPI rank per
node, a block size of 512, a look-ahead \texttt{depth} of 1.
We used the \texttt{increasing-2-ring} broadcast with the \texttt{Crout} panel
factorization algorithms and \(\texttt{P}\times\texttt{Q}=8\times32\) and we
tested matrix sizes ranging from 100,000 to 500,000. Let us denote by
\(T(N,C_i,\gamma)\) the performance of HPL when factorizing a matrix of rank
\(N\) on cluster \(C_i\) with a temporal variability of \(\gamma\). The overhead
for this configuration is the ratio
\begin{equation*}
O(N,C_i,\gamma) = \frac{\mathbb{E}[T(N,C_i,\gamma)]}{T(N,C_i,0)}-1.
\end{equation*}
\begin{figure}[!t]
    \centering
    \includegraphics[width=\linewidth]{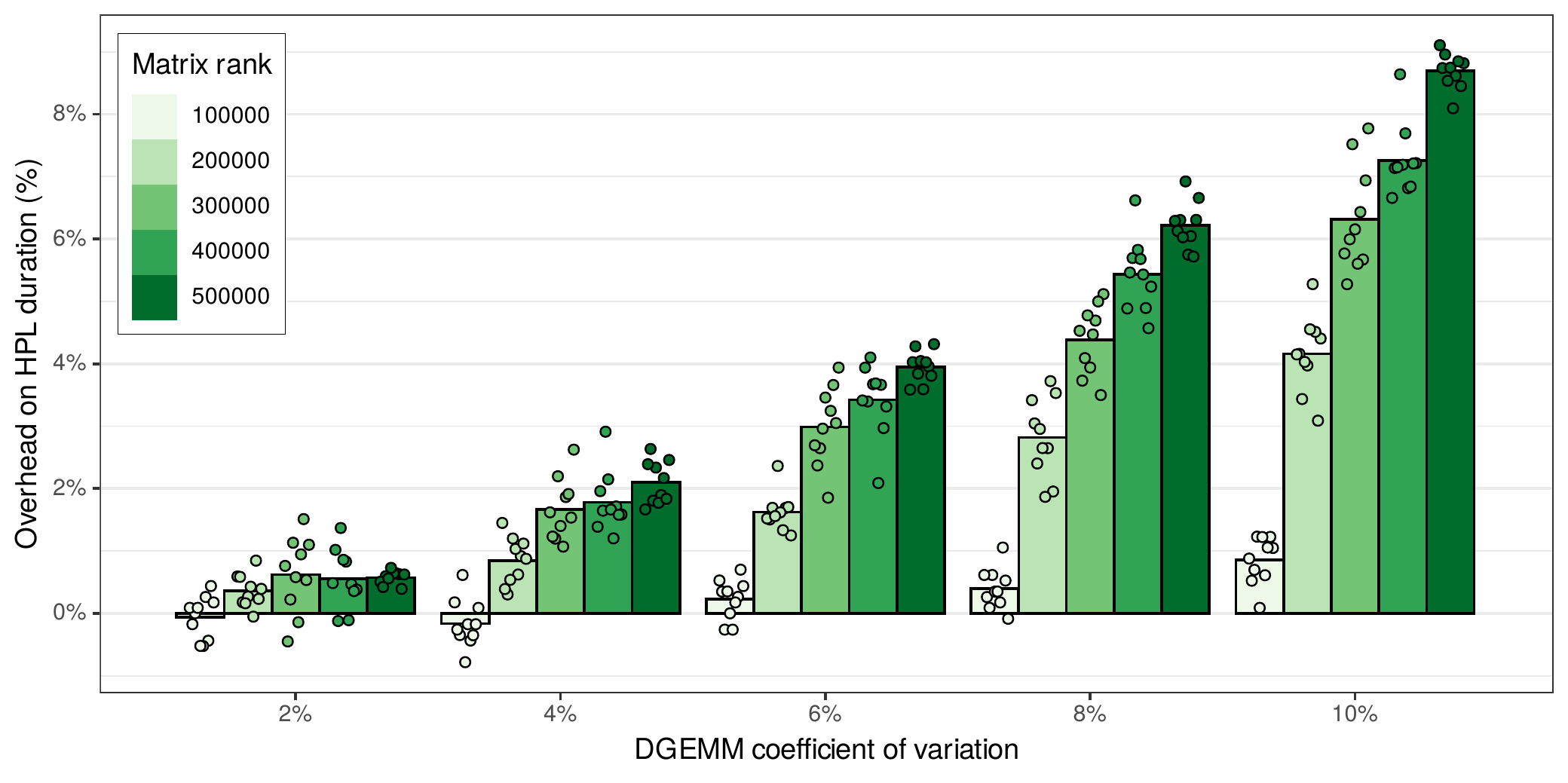}

    \caption{The overhead on HPL duration appears to be linear in \texttt{dgemm} temporal variability. Although it is negligible for small matrices, it severely inflates for larger matrices.}\vspace{-1em}
    \label{fig:whatif_variability}
    \labspace
\end{figure}
Each bubble in Figure\ref{fig:whatif_variability} represents one such
overhead. For any \(\gamma\), this overhead appears to be negligible for
small matrices and to increase and flatten when \(N\) grows large. In
most TOP5000 qualification runs, the matrix is made as large as
possible and the overhead would thus appear to grow roughly linearly
with \(\gamma\). On a new cluster, a simple statistical evaluation of the
nodes' performance using the model of
Section\ref{sec:whatif.model} would thus be a good first diagnosis of
whether trying to decreasing temporal variability is a promising
tuning target or not.

\subsection{Influence of Spatial Variability}
\label{sec:org427fd2e}
\label{sec:whatif.spatial_variability}
Although we showed in Section\ref{sec:validation.single} that
temporal variability could account for about 9\% of performance,
spatial variability was even more important as it was
responsible for 22\% of overhead compared to a fully homogeneous
cluster. In practice, the replacement of a few nodes may be possible
but such spatial variability is expected and 
common\cite{rountree_15} and a workaround would have to be found. A
common approach consists in dropping out a few of the slowest
nodes. Indeed, since the matrix is evenly divided between the nodes,
the computation inevitably progresses at the speed of the slowest
node. However, removing the slowest nodes also decreases the overall
processing capability and impacts the virtual topology's geometry
(the \texttt{P} and \texttt{Q} parameters of HPL). Such adjustment is often done by
trial and error and is all the more tricky as temporal
variability and uncertainty from real experiments come into play. In
this section, we show how such a subtle trade-off can be studied in
simulation.

\begin{figure}[pt]
    \centering
    \includegraphics[width=\linewidth]{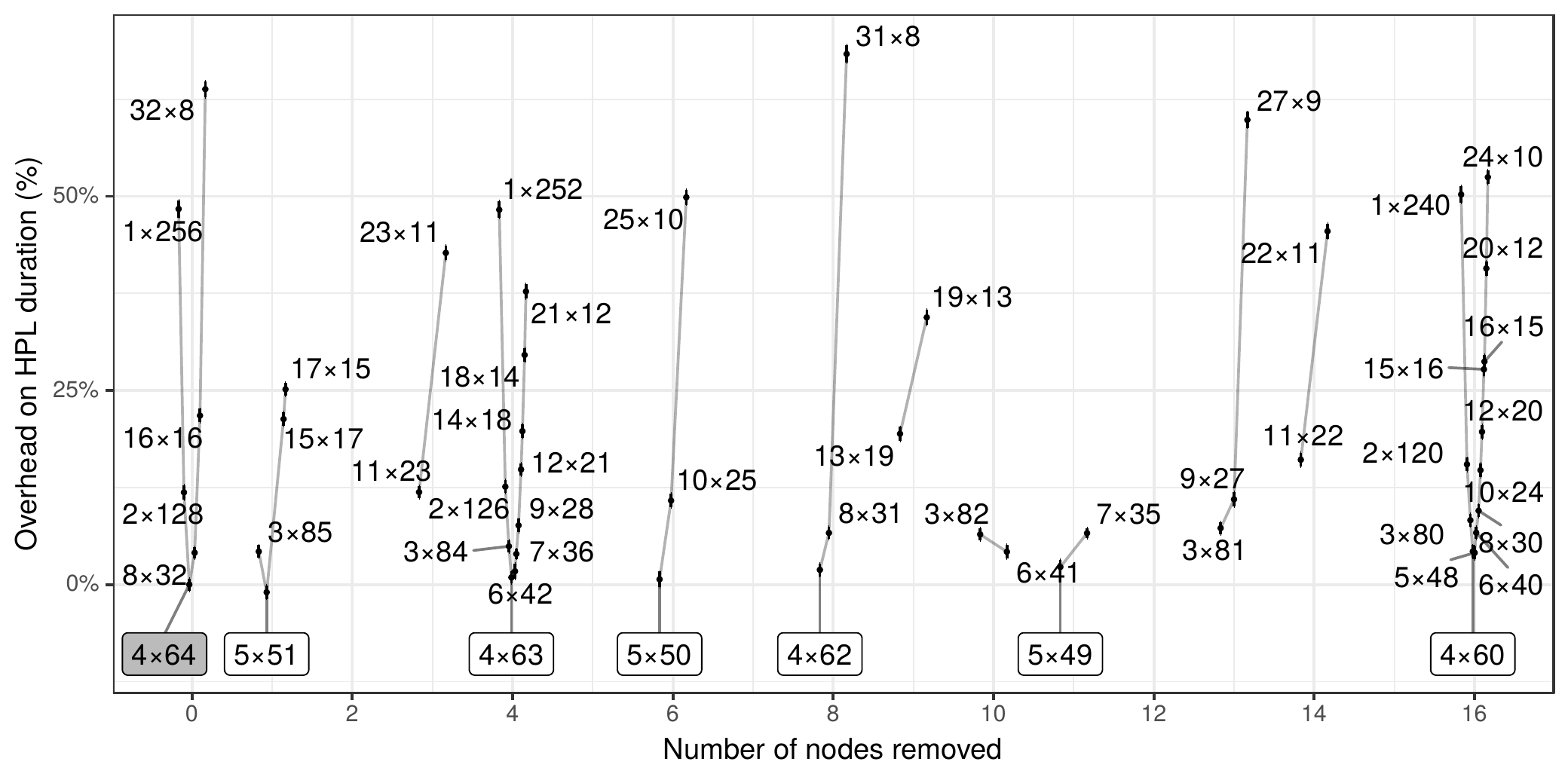}

    \caption{Influence of the number of nodes and of the geometry of the virtual topology on the performance of
      HPL\@: $\mathtt{P}\times\mathtt{Q}$
      configurations with a small $\mathtt{P}$ perform significantly
      better then those with a larger $\mathtt{P}$.}\vspace{-1em}
    \label{fig:whatif_removing_nodes}
    \labspace
\end{figure}

\begin{figure}[!t]
    \centering
    \includegraphics[width=.9\linewidth]{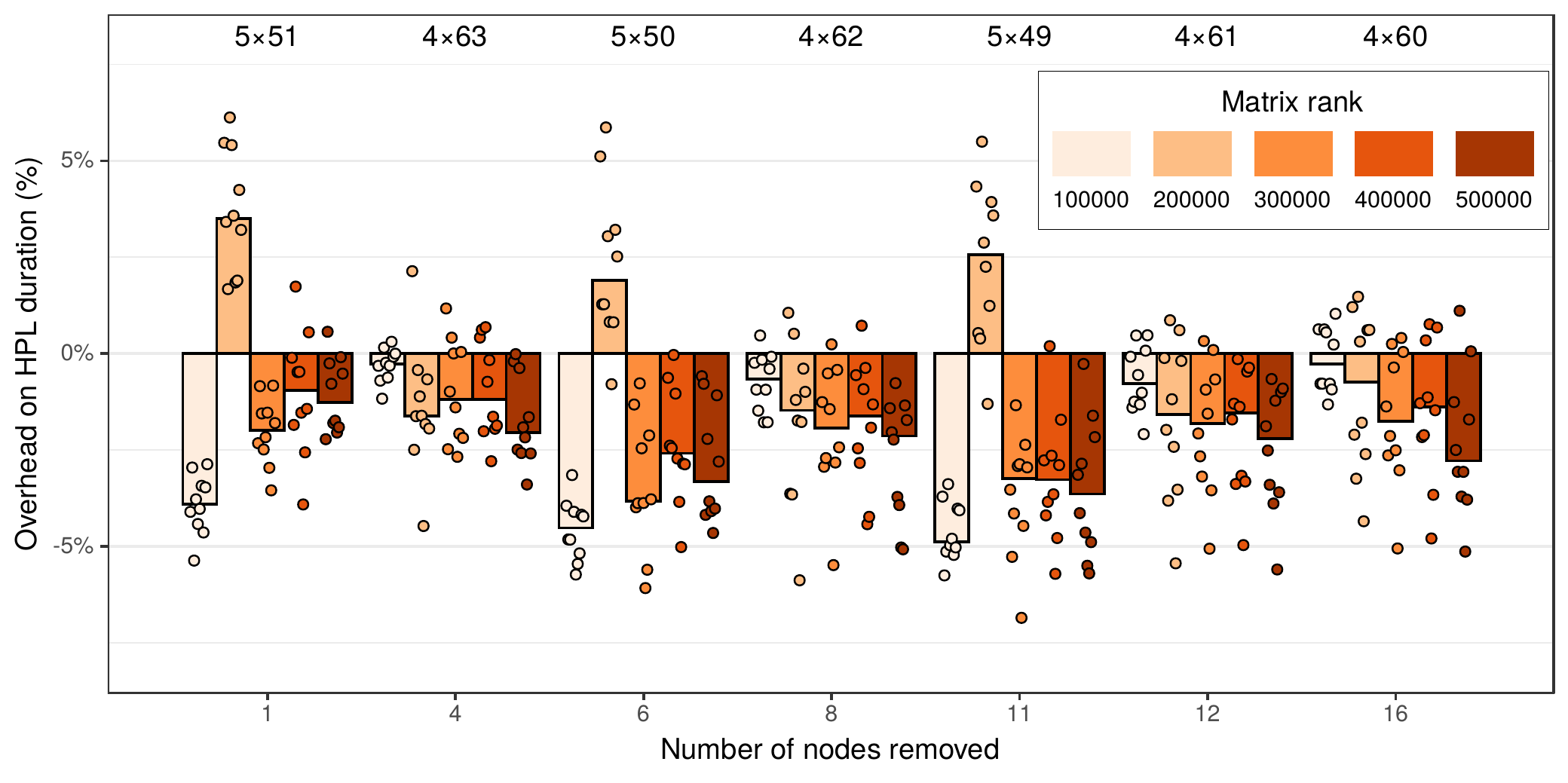}

    \caption{Influence of node removal on performance in a stronger
      heterogeneity scenario (extrapolation of our test cluster when
      it had a cooling problem on 4 of its nodes). Removing 6 to 12
      nodes our of 256 nodes may bring substantial improvement and
      such optimization would therefore be worth
      investigating.}\vspace{-1em}
    \label{fig:whatif_removing_nodes_heterogeneous}
    \labspace
\end{figure}

Using the model from Section\ref{sec:whatif.model}, we generate 10
mildly heterogeneous 256 node clusters (i.e., where nodes are similar
to the ones of our cluster when operating in the normal state as in
Figure\ref{fig:whatif_calibration}) and we study the performance
obtained when removing 1 to 16 of the slowest nodes. When removing
nodes, the geometry should be adjusted depending on how the number of
remaining nodes decomposes in prime factors. As observed in
Figure\ref{fig:validation_geometry}, having \texttt{P} \(\approx\) \texttt{Q} is generally a
good idea to reduce the total amount of communication. However
it may be counter-productive for a given broadcast or swap algorithm
that serializes communications. Figure\ref{fig:whatif_removing_nodes}
shows the overhead for a matrix of rank
250,000 compared to the best performance obtained using the whole
cluster. We group the different \(\texttt{P}\times\texttt{Q}\)
decompositions and order them by increasing \texttt{P}. Again, we use the
\texttt{2-Ring} and \texttt{Binary-exch} algorithms, which are among the best
configurations according to the study of
Section\ref{sec:validation.factorial}.  Each configuration is summarized through
the average overhead over the 10 clusters and errorbars represent a 95\%
confidence interval.
It appears that the \(4\times64\)
geometry now achieves the best trade-off between the total amount of
communications and how well they overlap with each other. The
optimal configuration for each number of nodes is boxed in
Figure\ref{fig:whatif_removing_nodes}. It reveals
that there is not much to gain, probably because of the mild spatial
heterogeneity of our cluster, but that optimizing the virtual topology
is particularly important.
\begin{figure}[!t]
    \centering
    \includegraphics[width=.9\linewidth]{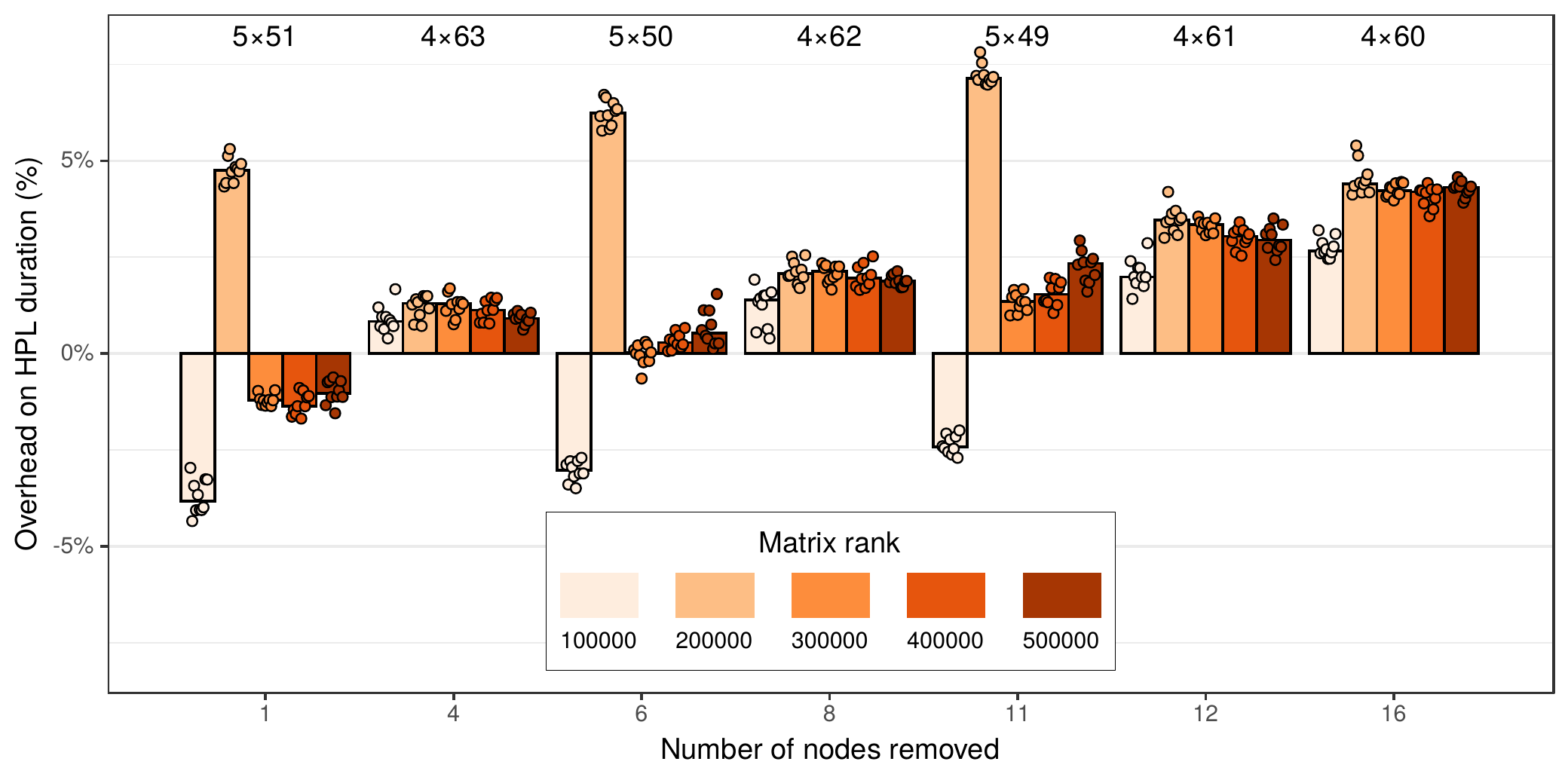}

    \caption{Influence of node removal on performance while
      taking into account the matrix rank. Due to the mild
      heterogeneity of these scenarios, evicting nodes brings no benefit.}\vspace{-1em}
    \label{fig:whatif_removing_nodes_2}
    \labspace
\end{figure}
Figure\ref{fig:whatif_removing_nodes_2} investigates how this overhead
for the best geometry and node selection also depends on the matrix
rank. It appears that in this scenario, except for very small
matrices, removing nodes cannot help improving performance. Note that
the overhead for \(5\times\texttt{Q}\) configurations with a matrix rank
200,000 appears to behave differently from what happens for other
matrix sizes. This surprising effect probably arises from a subtle
combination of matrix size and virtual topology. We could indeed
observe on our cluster that such configurations had a
weakly but significantly worse performance than the other
configurations. Such interaction also explains why designing a
faithful analytical model of HPL is so difficult and why a full
simulation of the whole application is generally required.
Although absolute performance should be taken with a grain of salt
when studying such subtle effects, they are easily overlooked when conducting
real experiments. In this particular small scale mild heterogeneity
scenario, there is thus no gain in removing nodes but, as illustrated in
Figure\ref{fig:whatif_removing_nodes_heterogeneous} where we used a multimodal
spatial heterogeneity (as in Figure\ref{fig:whatif_slow}), this may be a
relevant approach.  This sensibility analysis shows how, for a given
supercomputer, a simple statistical evaluation of the spatial heterogeneity
allows evaluating whether spatial variability is a promising tuning target or
not.

\subsection{Influence of the Physical Topology}
\label{sec:org641fd43}
\label{sec:whatif.topology}
\begin{figure}[pt]
    \centering
    \includegraphics[width=.9\linewidth]{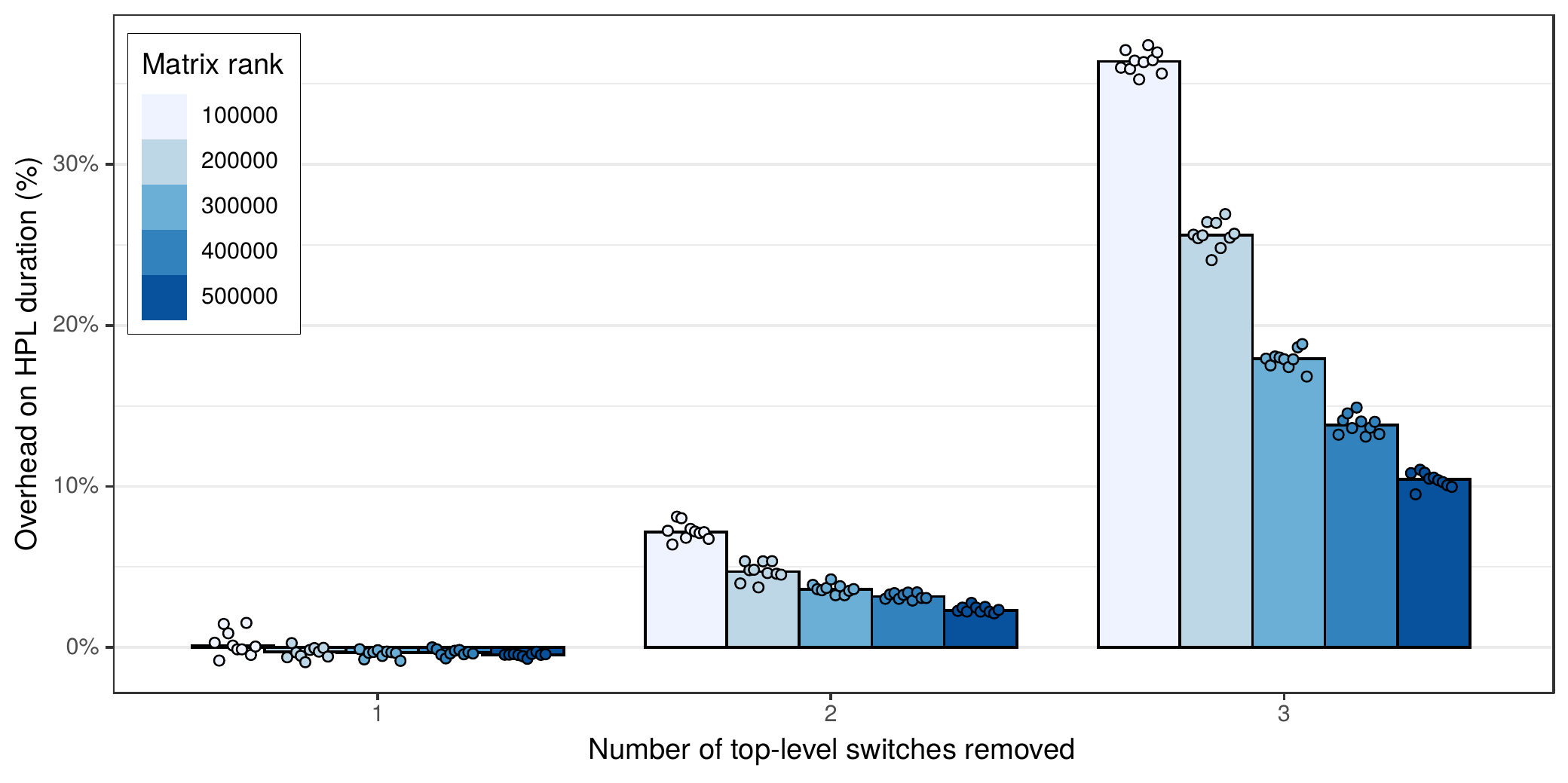}

    \caption{Influence of the physical topology on the overall performance. Up to 2 of the top-level switches can be removed without significantly hurting performances for large matrices. Beyond this point, communications become the main performance bottleneck.}\vspace{-1em}
    \label{fig:whatif_removing_switches}
    \labspace
\end{figure}

Finally, since virtual topology and communications appear to 
significantly influence the overall performance, one may wonder how
much the physical topology influences the performance. Indeed, several
recent articles\cite{tapered_fat_tree_16,tapered_fat_tree_19} report
that interconnect networks are often
oversized compared to the actual need of applications and that turning
off some switches could sometimes go completely unnoticed by
end-users. In this section we consider ten 256 node clusters with variable node
performance (as in Figure\ref{fig:whatif})
interconnected by a 2-level fat-tree and quantify by how much performance
degrades when the top-tier switches are gradually
deactivated. More formally, we use a \texttt{(2;32,8;1,N;1,8)} fat-tree with \(\texttt{N}\in\{1,2,3,4\}\).
Figure\ref{fig:whatif_removing_switches} depicts this
degradation as a function of matrix size. As one could expect, 
the impact is more significant for smaller matrix sizes (where
the execution is more network bound). Although removing one
switch leads to absolutely no visible performance loss, removing two
or three switches can have a dramatic effect. Again, such degradation
depends on the broadcast and swap algorithms and may be slightly
mitigated. To the best of our knowledge, it is the first time such
sensibility analysis is conducted faithfully. Generating
random node configurations allows avoiding potential bias, in
particular against perfectly homogeneous scenarios. We believe such
a tool can be quite useful in the earlier steps of a supercomputer
design when performing capacity planning to adjust the network
capacity to a given cost and power envelope.
\section{Discussion}
\label{sec:orgf490713}
\label{sec:relwork}

Analytical models are often used for estimating application performance. For
instance, Aspen\cite{aspen} is a domain specific language for specifying
formally the characteristics of kernels and combining them to model a whole
application. However, such an approach can only provide rough estimates, subtle
phenomena like network contention cannot be captured by these models.
Another approach for predicting the performance of applications like
HPL consists in statistical modeling the application as a whole\cite{hpl_prediction}. By
running the application several times with small and medium problem sizes (of a
few iterations of large problem sizes) and using simple linear regressions, it
is possible to predict its makespan for larger sizes with an error of only a few
percent and a relatively low cost. Unfortunately, the predictions are limited
to the same application configuration and studying the influence of the number
of rows and columns of the virtual grid, or the broadcast algorithms requires
a new model and new (costly) runs using the whole target machine. Our
attempts to build a black-box analytical model (involving,
polynomials, inverse, and logarithms of \texttt{P} and \texttt{Q}) of HPL from a limited
set of observations always failed to provide a faithful model with
decent prediction and extrapolation capabilities. Furthermore,
this approach does not allow studying \emph{what-if} scenarios (\eg to evaluate what
would happen if the network bandwidth was increased or if node heterogeneity was
decreased) that go beyond parameter tuning.

Simulation provides the details and flexibility missing to such a black-box
modeling approach. Performance prediction of MPI applications through simulation
has been widely studied over the last decades but two approaches can be
distinguished in the literature: offline and online simulation.

With the most common approach, \emph{offline simulation}, a trace of the application is
first obtained on a real platform. This trace comprises sequences of MPI
operations and CPU bursts and is given as an input to a simulator that
implements performance models for the CPUs and the network to derive
predictions. Researchers interested in finding out how their application reacts
to changes to the underlying platform can replay the trace on commodity hardware
at will with different platform models.  Most HPC simulators available today,
notably BigSim\cite{bigsim_04}, Dimemas\cite{dimemas} and CODES\cite{CODES},
rely on this approach.  The main limitation of this approach comes from the
trace acquisition requirement. Not only is a large machine required but the
compressed trace of a few iterations (out of several thousands) of HPL typically
reaches a few hundred MB, making this approach quickly
impractical\cite{suter}. Worse, tracing an application provides only information
about its behavior of a specific run: slight modifications (\eg to
communication patterns) may make the trace inaccurate. The behavior of simple
applications (\eg \texttt{stencil}) can be extrapolated from small-scale
traces\cite{scalaextrap,pmac_lspp13} but this fails if the execution is
non-deterministic, \eg whenever the application relies on non-blocking
communication patterns, which is, unfortunately, the case for
HPL\@.

The second approach discussed in the literature is \emph{online simulation}.
Here, the application is executed (emulated) on top of a platform simulator
that determines when each process is run. This
approach allows researchers to study directly the behavior of MPI
applications but only a few recent simulators such as SST
Macro\cite{sstmacro}, SimGrid/SMPI\cite{simgrid} and the closed-source
xSim\cite{xsim} support it. To the best of our knowledge, only SST
Macro and SimGrid/SMPI are mature enough to faithfully emulate HPL. In
this work, we decided to rely on SimGrid as its performance models and
its emulation capabilities are quite solid but the work we present would a
priori also be possible with SST.  The SST simulator has also been used for
uncertainty quantification (UQ)\cite{sst_uq}, but with a different purpose than
what we did in Section\ref{sec:whatif}.
Note that the HPL
emulation we describe in Section\ref{sec:em} should not be confused
with the application skeletonization\cite{sst_skeleton} commonly used
with SST and more recently introduced in CODES. Skeletons are code
extractions of the most important parts of a complex application
whereas we only modify a few dozens of lines of HPL before emulating
it with SMPI.  Some researchers from Intel unaware of our recent work
recently applied the same methodology as the one we proposed
in\cite{cornebize:hal-02096571} to both Intel HPL and OpenHPL in the
closed-source CoFluent simulator\cite{intel_20}. To the best of our
knowledge, their work reports two faithful predictions for two
large-scale supercomputers but without investigating at all the impact
of variability, heterogeneity, nor of communications as we do in this
article. Finally, it is important to understand that the approach we
propose is intended to help studies at the whole machine
and application level, not the influence of microarchitectural details as
intended by gem5\cite{lowepower2020gem5} or MUSA\cite{musa_16}.
\section{Conclusion}
\label{sec:org55885b3}
\label{sec:cl}

HPC application developers implement many elaborate algorithmic strategies whose
impact on performance is often dependent on both the input workload
and the target platform. This structure makes it very difficult to
model and accurately forecast the overall application performance, and
many HPC application developers and users are often left with no other
option but to study and tune their applications at scale, which can
be very time- and resource-consuming. We believe that being capable of
precisely predicting an application's performance on a given platform
is useful for application developers and users and will become
invaluable in the future as it can, for example, help computing centers
with deciding which one of the envisioned technologies for a new
machine would work best for a given application or if an upgrade of
the current machine should be considered.

Simulation is an effective approach in this context and SimGrid/SMPI
has previously been successfully validated in several small-scale
studies with simple HPC benchmarks\cite{smpi,heinrich:hal-01523608}.
In an earlier work\cite{cornebize:hal-02096571}, we have explained how
SMPI could be used to efficiently emulate HPL. The proposed approach
only requires minimal code modifications and applies
to any application whose behavior does not strongly depend on
data-dependent intermediate computation results. Although HPL is not a
\emph{real} application, it is quite optimized from an algorithmic point of
view and its behavior can be controlled through 6 different
parameters (granularity, geometry of the virtual topology,
broadcast/swapping/factorization algorithm, and the number of
concurrent iterations). HPL features classical optimization techniques
such as heavily relying on \texttt{MPI\_Iprobe} to overlap communication with
computations, making it particularly challenging both in terms of
tuning and simulation.

In this article (Section\ref{sec:smpi} and\ref{sec:validation}), we
present an extensive validation study which covers the whole parameter
space of HPL. Our study emphasizes the importance of carefully
modeling (1) the platform heterogeneity (not all nodes have exactly
the same performance), (2) the short-term temporal variability (e.g.,
system noise) for compute kernels as it may propagate in
communication patterns, and (3) the complexity of MPI (performance often
wildly differs between small and large messages and between intra-node
and extra-node communications). We show that disregarding any of these
aspects may lead to wildly inaccurate predictions even on an
application as regular as HPL. By building on a few well-identified
micro-benchmarks of the BLAS and MPI, we show that these aspects
can be well modeled, which allows us to systematically predict the
overall performance of HPL within a few percent. Our experimental
results span over two years and we report situations (in
Section\ref{sec:methodology.timeframe}
and\ref{sec:validation.geometry}) where the simulation helped us to
identify performance regression or anomalies incurred by the platform
when the prediction did not match the real experiments.

We show (in Section\ref{sec:validation}) how this faithful surrogate
can be used to evaluate the significance of application parameters and
tune them accordingly solely through simulations.  We also propose a
generative model for the compute nodes' performance that can
easily be fit from daily measurements and used to produce synthetic
platforms similar to the ones at hand. We show (in
Section\ref{sec:whatif}) how this model, which allows us to easily
control temporal and spatial variability, can feed our simulations
to assess the impact of variability on the performance of the
application or of mitigation strategies (e.g., the eviction of the
slower nodes). Likewise, the simulation allows to easily assess the
influence of the physical network on the overall performance. Most of
these \emph{what-if} studies would be particularly difficult to conduct
through real experiments because of the difficulty to finely control
the platform. This is to the best of our knowledge one of the first
sensitivity analyses of a real HPC code accounting for platform
uncertainty.

As future work, building on the effort of SimGrid developers on
supporting the emulation of a wide variety of applications with
SMPI\cite{smpi_proxy_apps}, we also intend to conduct similar studies
with other HPC benchmarks (\eg HPCG\cite{HPCG} or HPGMG\cite{HPGMG}),
real applications (\eg BigDFT\cite{bigdft}) and larger
infrastructures. As explained in this article, a good model of
compute kernels and the MPI library is essential. Thereby, the main
challenge for systematic use of our simulation technique now lies in the
automation of measurements through well-designed experiments and 
the automatic detection of when the envisioned models miss essential
characteristics of the platform (multi-modal behaviors,
heteroscedasticity, discontinuities,\dots{}). We intend to provide
a fully automatic calibration procedure for MPI as well as for every BLAS function, which
would allow us to effortlessly predict the performance of many
applications by simply linking against a BLAS-replacement library.

\def\subsection#1{\osubsection*{#1}}
\subsection{Acknowledgments}
\label{sec:org91e6663}

Experiments presented in this paper were carried out using the Grid'5000
testbed, supported by a scientific interest group hosted by Inria and including
CNRS, RENATER, and several Universities as well as other organizations (see
\url{https://www.grid5000.fr}).
Last, we warmly thank the SimGrid developers for their help in
integrating our contributions and for their early feedback on
this work before submission.

This research did not receive any specific grant from funding agencies in the public, commercial, or not-for-profit sectors.
\pagebreak

\bibliographystyle{elsarticle-num}
\bibliography{refs}
\end{document}